\documentclass[12pt,english,a4paper]{ar-1col}
\usepackage[T1]{fontenc}
\usepackage[latin9]{inputenc}
\usepackage{xcolor}
\usepackage{pdfcolmk}
\usepackage{babel}
\usepackage{amstext}
\usepackage{amssymb}
\usepackage{graphicx}
\usepackage{setspace}
\PassOptionsToPackage{normalem}{ulem}
\usepackage{ulem}
\usepackage[unicode=true,
 bookmarks=true,bookmarksnumbered=false,bookmarksopen=false,
 breaklinks=false,pdfborder={0 0 0},pdfborderstyle={},backref=false,colorlinks=false]
 {hyperref}

\makeatletter

\providecolor{lyxadded}{rgb}{0,0,1}
\providecolor{lyxdeleted}{rgb}{1,0,0}
\DeclareRobustCommand{\lyxsout}[1]{\ifx\\#1\else\sout{#1}\fi}

\usepackage{babel}
\makeatother

\begin{document}
\jname{Ann. Rev. Nucl. Part. Sci.} 
\jvol{68} 
\jyear{2018} 
\doi{10.1146/TBD} 

\title{Heavy Ion Collisions: \\The Big Picture, and \\the Big Questions}

\author{Wit Busza,$^1$ Krishna Rajagopal,$^{1,2}$ and\\ Wilke van der Schee$^{2, 3}$ 
\affil{$^1$Laboratory for Nuclear Science and Department of Physics, MIT, 77 Massachusetts Avenue, Cambridge, MA 02139, USA}  
\affil{$^2$Center for Theoretical Physics, Massachusetts Institute of Technology, 77 Massachusetts Avenue, Cambridge, MA 02139, USA}  
\affil{$^3$Institute for Theoretical Physics and Center for Extreme Matter and Emergent Phenomena, Utrecht University, Leuvenlaan 4, 3584 CE Utrecht, The Netherlands} 
}

\begin{abstract}
Heavy ion collisions quickly form a droplet of quark-gluon plasma (QGP) with a remarkably small viscosity. We give an accessible introduction to how to study this smallest and hottest droplet of liquid made on earth and why it is so interesting. The physics of heavy ions ranges from highly energetic quarks and gluons described by perturbative QCD to a bath of strongly interacting gluons at lower energy scales. These gluons quickly thermalize and form QGP, while the energetic partons traverse this plasma and end in a shower of particles called jets. Analyzing the final particles in a variety of different ways allows us to study the properties of QGP and the complex dynamics of multi-scale processes in QCD which govern its formation and evolution, providing what is perhaps the simplest form of complex quantum matter that we know of. Much remains to be understood, and throughout the review big open questions will be encountered.
\end{abstract}


\maketitle
\tableofcontents

\section{Introduction \label{sec:INTRODUCTION}}

In the past 50 years, as beams of ultra relativistic protons and nuclei
have become available, the collisions of protons with nuclei and nuclei
with nuclei, at higher and higher relative velocities (or total collision
energy), have been studied in greater and greater detail. This review
provides some answers to the questions: ``Why do such studies?''
and ``What have we learned from them so far?'' and ``What are the
big questions that they may illuminate in future?''.

We start with a qualitative description, in the center of mass frame
(the ``lab frame'' at a collider), of the sequence of events that
occur when two ultra relativistic nuclei collide, head on. This picture
follows from the observed phenomenology (summarized in Section \ref{sec:Phenomenology-of-heavy}),
relativity, and our understanding of the workings of QCD.\footnote{For the reader interested in a more comprehensive introduction into
heavy ion collisions we refer to the books \cite{Yagi:2005yb,Florkowski:2010zz,Wang:2016opj}.
Note also that because we are required to limit the number of citations,
throughout this review where possible we will cite articles in which
more citations can be found. }

Each incident nucleus is a Lorentz contracted disc. For large nuclei
such as Pb or Au, the diameter of the disc is about 14 fm (femtometer,
or Fermi) and its thickness is about $14/\gamma$ fm, where, at the
highest beam energies attainable at RHIC and LHC, the relativistic
$\gamma$ factors are approximately 100 and 2500 respectively, corresponding
to beam rapidities of $y=5.3$ and $8.5$. \begin{marginnote}[] \entry{Momentum rapidity $y$}{$\cosh (y) \equiv \gamma$, with $\gamma = 1/\sqrt{1-v_z^2}$, with $v_z$ the velocity along the beam direction in units of the speed of light.} \entry{Space-time rapidity $y_s$}{$\tanh (y_s) \equiv z/t$, with $z$ and $t$ space-time coordinates centered at the collision. $y_s$ cannot be measured experimentally; however, in a boost invariant flow in which $v_z=z/t$, $y_s=y$.}
\end{marginnote}Each disc includes many colored quarks and antiquarks, with three
more quarks than antiquarks per nucleon in the incident nuclei and
with $q\bar{q}$ pairs coming from quantum fluctuations in the initial
state wave functions that are ``almost real'', as a consequence
of time dilation. These quarks and antiquarks are, in turn, sources
of strong, almost completely transverse, color fields and corresponding
field quanta, the gluons which also carry color.

The area density of quarks, antiquarks and gluons, partons for short
in the language of Feynman, increases with the velocity of the nuclei.
It is not uniform across the area of the disc, and fluctuates from
nucleus to nucleus. The spatial variation of the partons primarily
reflects the instantaneous distribution of the nucleons inside the
nuclei and of the partons inside the nucleons. Over all, the incident
nuclei are highly complex systems of partons with a longitudinal momentum
distribution (referred to as a structure function) that is close to
being a superposition of that in the individual nucleons but with
small modifications coming from the proximity, and motion, of nucleons
in nuclei.

When the two discs, each a tiny fraction of a fm thick, overlap or
collide, most of the incident partons lose some energy but are not
kicked by any large angle. Most of these interactions are ``soft'',
meaning that they involve little transverse momentum transfer. These
strong interactions can be described in terms of interacting fields
or slabs of energy. In the language of fields and particles, as the
two discs of strongly interacting transverse color fields and associated
color charges collide, some color charge exchange occurs between the
discs, and longitudinal color fields are produced, which fill the
space between the receding two discs, reducing the energy in the discs
themselves, and then gradually decay into $q\bar{q}$ pairs and gluons.
A small fraction of the incident partons suffer hard perturbative
interactions as the discs overlap initially which, as we will discuss
later, lead to a relatively improbable but very important production
of particles with high transverse momentum. 

\begin{figure}
\begin{centering}
\includegraphics[width=1.2\textwidth]{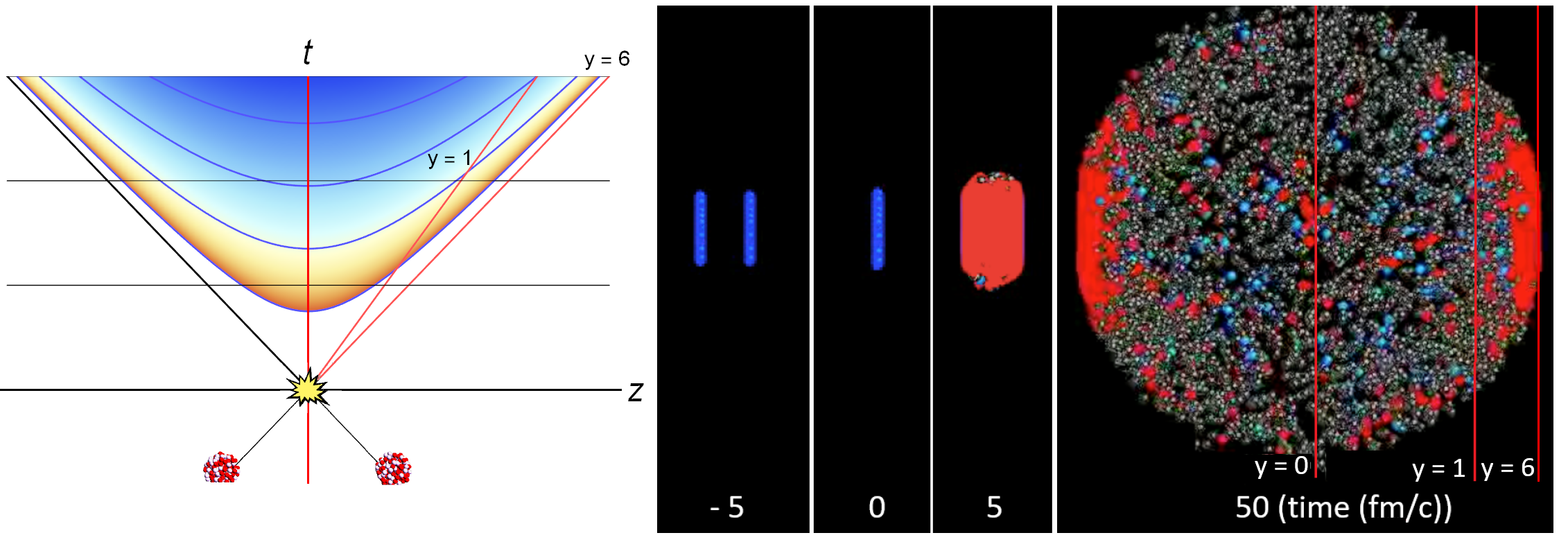}
\par\end{centering}
\caption{\textbf{(left) }Space-time picture of a heavy ion collision, whereby
the color gives an indication of the temperature of the plasma formed.
Dynamics takes place as a function of proper time (blue curves), which
is why plasma forms later at higher rapidities. \textbf{(right) }Snapshots
of a central 2.76 TeV PbPb collision at different times (different
horizontal slices of the space-time picture on the left) with hadrons
(blue and grey spheres) as well as QGP (red). In both figures, at
a given time the hottest regions can be found at high rapidity close
to the outgoing remnants of the nuclei and the red lines indicate
the approximate longitudinal location of particles with rapidity $y=0$,
$y=1$, and $y=6$. (Figs. adapted from \cite{vanderSchee:2014qwa,HICmovie}.)\label{fig:movie}}
\end{figure}

In high energy heavy ion interactions, the maximum energy density
occurs just as the two highly Lorentz contracted nuclei collide. Clearly
this system is very far from equilibrium, and its very high energy
density is really just a consequence of Lorentz contraction. It is
much more interesting to ask what we can say in a generic way about
the average energy density say 1 fm/c after the collision, by which
time the two discs are 2 fm apart. The expanding high energy density
system produced around the midpoint between the two discs, where the
collision occurred, has an energy density at that time that is still
far in excess of $500\text{ MeV/fm}^{3}$, the energy density inside
a typical hadron. A rough estimate can be obtained from the available
data for head-on LHC collisions with $\sqrt{s_{NN}}=2.76$ TeV\begin{marginnote}[] \entry{$\sqrt{s_{NN}}$}{is the total collision energy per nucleon-nucleon pair in the center of mass frame. %For example for Pb-Pb ($A=208$) at 5.02 TeV the total energy of one nucleus is 522 TeV.
}
\end{marginnote} (corresponding to $\gamma=1400$ and $y=8.0$) by noting that the
total transverse energy in particles with pseudorapidity between -0.5
and 0.5\begin{marginnote}[] \entry{Pseudorapidity}{$\eta\equiv - \log \left[ \tan \left( \theta/2 \right) \right]$, where $\theta$ is the polar angle in momentum space relative to the beam direction. $\eta$ is a standard proxy for rapidity $y$ because $\eta=y$ for massless particles.
}
\end{marginnote}(so longitudinal velocity $-0.46<v<0.46$) is measured to be $1.65\pm0.1$
TeV \cite{Collaboration:2011rta}, meaning that the average energy
density 1 fm/c after the collision is greater than $1.65\text{ TeV}/(\pi(7\text{ fm})^{2}(0.92\text{ fm}))=12\text{ GeV/fm\ensuremath{^{3}}}$,
about twenty times the energy density of a hadron. The entropy produced
in these collisions is also enormous; to get a sense of this note
that before the collision the entropy of the two incident nuclei is
essentially zero whereas the final state after the collision can contain
as many as 30,000 particles, and hence has a very large entropy. We
shall return to this later, and in particular we shall see that most
of this entropy is produced quickly, in the initial moments after
the collision. To get a further sense of the magnitude of the average
energy density 1 fm/c after the collision, note that, as we shall
see in Section \ref{sec:Phenomenology-of-heavy}, lattice calculations
of QCD thermodynamics show that matter in thermal equilibrium at a
temperature of 300 MeV has an energy density $\approx12\,T^{4}=12.7\,\text{GeV}/\text{fm}^{3}$.
Thus, the quarks and gluons produced in the collision cannot be described
as a collection of distinct individual hadrons. Nevertheless, the
quarks and gluons in this high energy density matter are far from
independent. They are so strongly coupled to each other that they
form a collective medium that expands and flows as a relativistic
hydrodynamic fluid with a remarkably low viscosity to entropy density
ratio $\eta/s\approx1/4\pi$ \cite{Heinz:2013th,Romatschke:2017ejr},
in units with $\hbar=k_{B}=1$, within a time that can be shorter
than or of order 1 fm/c in the rest frame of the fluid. This form
of matter has been named Quark-Gluon Plasma, or QGP for short. Even
if the transverse velocity of the fluid is small initially, say 1
fm/c after the collision, the pressure-driven hydrodynamic expansion
rapidly builds up transverse velocities of order half the speed of
light. As the discs recede from each other and the QGP produced between
them is expanding and cooling, at the same time new QGP is continually
forming in the wake of each receding disc, see Fig. \ref{fig:movie}.
This happens because the quarks and gluons produced at high rapidity
are moving at almost the speed of light in one of the beam directions,
meaning that when enough time has passed in their frame for them to
form QGP a long time has passed in the lab frame, around $330$ fm/c
for rapidity $y=6.5$. Throughout this QGP production process, each
disc gradually loses energy as partons with higher and higher rapidity
separate from it and form QGP. In contrast, the occasional high transverse
momentum particles seen in some collisions are produced by large-angle
scattering at very early times, when the incident nuclei collide.

The process ends once QGP has formed at the rapidities where most
of the baryon number from the incident nuclei ends up, which is expected
to be about 2 units of rapidity less than that of the incident nuclei,
based upon measurements made in lower energy proton-nucleus collisions
\cite{Busza:1989px}. So, the discs lose about 85\% of their energy
while varying amounts of QGP form at varying rapidities over a range
that extends between $y=-6.5$ and $y=6.5$ in collisions at the LHC.
A good way to visualize the QGP production process described above
is to consider the production of each volume element of QGP in its
own local rest frame, where the two colliding nuclei have an asymmetric
rapidity and energy, and then boost this volume of QGP back to the
lab frame.

After production, each elemental volume of QGP expands in all directions.
Looked at overall, the droplet of fluid flows hydrodynamically, as
its initial high pressure drives fluid motion, expansion, and consequent
cooling. This picture holds until the energy density at a given location
in the fluid drops below that within an individual hadron, at which
point the fluid falls apart into a mist of hadrons that scatter off
each other a few times and then stream away freely. This mechanism
of particle production, via an intermediate epoch during which a hydrodynamic
fluid forms and expands, is quite different from the current understanding
of particle production in elementary collisions in which only a few
new particles are created.

Meanwhile, remnants of the original nuclei (excited nuclear matter
compressed by about a factor of 5-10 \cite{Busza:1983rj}) progress
in the forward and backward directions. This high baryon density system
then expands and hydrodynamizes, forming hot quark-rich QGP after
a time of order 1 fm/c in its own rest frame, corresponding to a time
of about $330$ fm/c in the lab frame. After further expansion, it
subsequently falls apart into hadrons. Unfortunately, none of the
LHC detectors are adequately instrumented around $y=6.5$, an almost
impossible task, meaning that the debris formed from this hot, high
baryon density, QGP has not yet been studied.

So far we have considered head-on (``central'') collisions. \begin{marginnote}[] \entry{Central and off-central}{Ions colliding head-on are called central collisions, whereas if the ions only partially overlap the collision is non-central or peripheral.} \end{marginnote}How
about non-central collisions? In the overlap region, the process is
the same as described above, except that the droplet of QGP is formed
with an initial approximately lenticular shape in the transverse plane.
In reality, because nuclei are made of individual nucleons the energy
density of the QGP that forms is lumpy in the transverse plane, making
it neither perfectly circular in head-on collisions nor perfectly
lenticular in non-central collisions. Deviations from circular symmetry
in the initial shape of the QGP, whether due to off-center collisions
or the lumpiness and fluctuations of the incident nuclei, result in
anisotropies in the pressure of the hydrodynamic fluid, which in turn
drive anisotropies in the expansion velocity and hence in the azimuthal
momentum distribution of the finally produced particles. 

In an off-center (non-central) collision, the parts of the incident
nuclei that do not collide are referred to as spectators. At very
early times, they create a magnetic field in the collision zone whose
possible effects are the subject of much discussion that we do not
have space to review \cite{Kharzeev:2015znc}. Later, they fragment
into excited nuclei and hadrons, moving with almost the full rapidity
of the incident nuclei. The extreme limit of an off-center collision
is one in which the nuclei themselves miss each other but the Lorentz-contracted
disc of electromagnetic fields around them do interact. These ultraperipheral
collisions give rise to copious $\gamma\gamma$ and $\gamma A$ interactions
\cite{Baltz:2007kq}, \begin{marginnote}[] \entry{Particle nomenclature}{Collisions are often described by shorthands $p$, $A$, $\gamma$, $\pi$ and $K$ for protons, heavy ions, photons, pions and kaons respectively}
\end{marginnote}which we shall not pursue except to note here that they dominate the
total nucleus-nucleus interaction cross-section.

Finally, a word about the hard collisions between two partons in the
incident nuclei. Such collisions, especially where particles with
very large transverse momenta (say, greater than a few tens of GeV/c)
are produced, are rare but very important. They lead to the production,
essentially from a point at the earliest times of the overall collision
process, of high-energy parton pairs and electroweak bosons. The high-energy
partons evolve, decay, radiate and finally produce a cone-shaped spray
or ``jet'' of hadrons and/or high-energy photons, leptons or heavy
$Q\bar{Q}$ pairs, all while traversing a region where QGP is in the
process of being produced and evolving. They thus contain a wealth
of information about the produced medium (in essence they ``X-ray''
the medium), and on how partons lose energy or disturb the medium
as they interact with it. We shall return to this in Section \ref{sec:Jets-in-quark-gluon}.

As is evident from the above description, collisions of ultrarelativistic
nuclei are complex, consisting of several distinct stages, each probing
different aspects of QCD. What makes them interesting is that the
regimes of QCD that they give us a means to explore are places where,
because of the strength of the QCD interactions, we would not even
have a zeroth order understanding of the properties of the matter
that QCD describes, let alone the dynamical phenomena, without having
seen what happens in these collisions. Heavy ion collisions are a
laboratory that is rich with unique ways to probe fundamental aspects
of QCD empirically, with some control over varying conditions. Our
description of a relativistic heavy ion collision brings us directly
to a set of questions: Do we have, at least qualitatively, an understanding
of all the stages of a heavy ion collision? Have any fundamentally
new phenomena been seen? Any unexplained phenomena? What new insights
have we obtained, or can we obtain, about the workings of QCD from
analysis of heavy ion collision data and corresponding theoretical
calculations? How, and how well, do we understand the initial stages
of the collision process, up to the creation of QGP? Which aspects
involve weakly coupled dynamics, and which strongly coupled? What
are the properties of QGP? From the study of jets and high momentum
particles, what have we learned about the properties of strongly interacting
matter, and about the dynamics of fast particles as they traverse
strongly coupled matter? What new insights have we obtained about
the formation of hadrons? Beginning in Section \ref{sec:Phenomenology-of-heavy},
we will attempt to answer some of these questions. But first, in the
next Section we shall expand our perspective.

\section{Why do we study ultrarelativistic heavy ion collisions?\label{sec:Why-do-we}}

The overarching answer with which we ended Section \ref{sec:INTRODUCTION},
albeit formulated as a suite of questions, is that studying ultrarelativistic
heavy ion collisions may give us a path to a more complete understanding
of how particles are produced in high energy collisions in QCD. This
is a fundamental question that in fact long predates QCD: Heisenberg
and Heitler wrestled with it in the 1930s and 1940s \cite{Heisenberg:1949kqa,Hamilton:1986fj},
Fermi and Landau did so in the 1940s and 1950s \cite{Fermi:1950jd,Landau:1953gs,Florkowski:2010zz},
and Feynman tried his hand in the 1960s \cite{Feynman:1969ej}. We
can now gain new purchase on these old questions by studying high
energy collisions in a new regime in which experimenters have new
knobs to dial including the size of each of the colliding nuclei,
(proxies for) the impact parameter, the final state multiplicity,
and more. 

In this Section, we shall formulate variants of the ``why are we
doing this?'' question that take us beyond the subject of ultrarelativistic
heavy ion collisions per se. Are there insights that we hope to gain
from studying these collisions that go beyond understanding the dynamics
of these collisions, or even of ultrarelativistic collisions in QCD
more generally? The affirmative answers to this question, which we
shall divide into three groups below, motivate much of the experimental
and theoretical efforts that we describe in this review.

\begin{figure}
\begin{centering}
\includegraphics[width=1.2\textwidth]{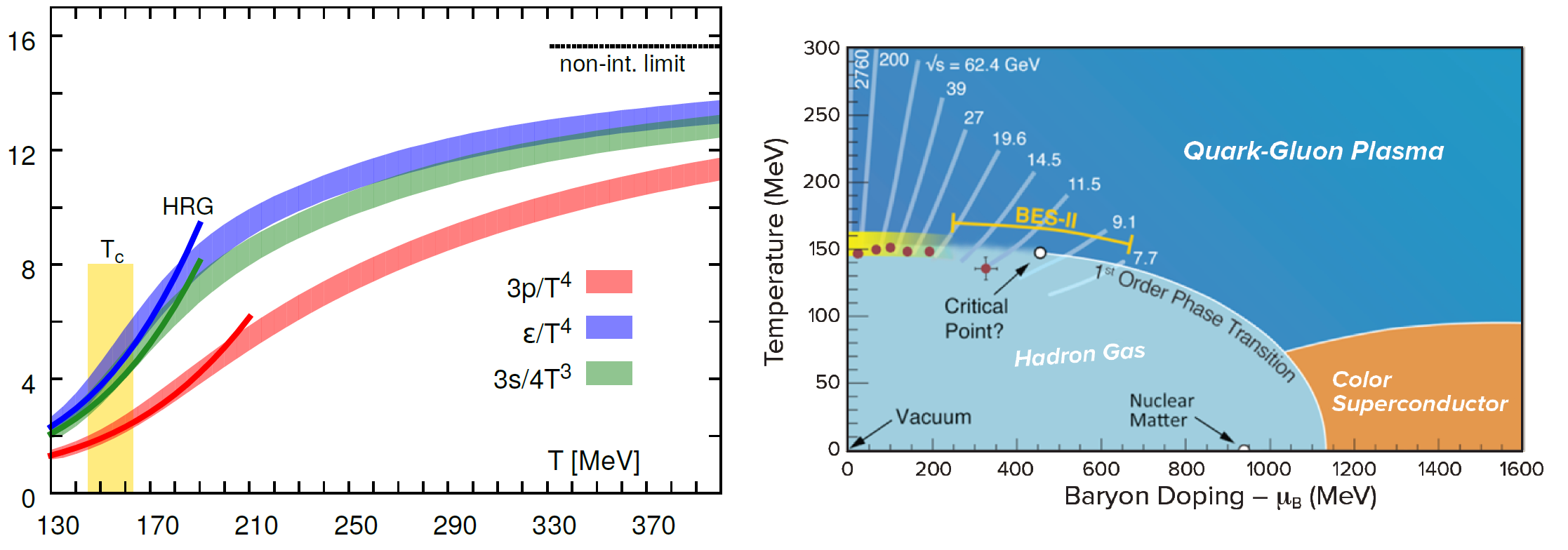}
\par\end{centering}
\caption{\textbf{(left) }Lattice QCD calculations of the pressure $p$, energy
density $\varepsilon$ and entropy density $s$ of hot QCD matter
in thermal equilibrium at temperature $T$ \cite{Borsanyi:2013bia,Bazavov:2014pvz}
show a continuous crossover around $T\sim150$ MeV, from a hadron
resonance gas (HRG) at lower temperatures to QGP at higher temperatures.
Because QCD is asymptotically free, thermodynamic quantities will
reach the Stefan-Boltzmann limit (weakly coupled quarks and gluons;
the non-interacting limit is marked in the figure) at extremely high
temperature. At the range shown, however, they are around 20\% below
their Stefan-Boltzmann values, which is consistent with simple estimates
for strongly coupled plasma based on holography \cite{Casalderrey:2014}.
The rise in $\varepsilon/T^{4}$ and $s/T^{3}$ seen in the figure
is a direct manifestation of the crossover from a hadron gas to QGP
which has more thermodynamic degrees of freedom because color is deconfined.
Using experimental data to constrain $\varepsilon/T^{4}$ remains
an outstanding challenge: comparing hydrodynamic calculations to various
experimental measurements gives us information about $\varepsilon$
versus time but, although some information about $T$ can be obtained
by analyzing measurements of photons, electrons and muons from heavy
ion collisions \cite{Paquet:2015lta}, at present $T$ cannot be determined
with sufficient accuracy to constrain $\varepsilon/T^{4}$ well enough
to see the rise in the number of degrees of freedom in QGP. \textbf{(right)}
This sketch illustrates our current understanding of the expected
features of the phase diagram of QCD as a function of temperature
and baryon doping, the excess of quarks over antiquarks, parametrized
by the chemical potential for baryon number $\mu_{B}$. The lattice
calculations in the left panel were done with $\mu_{B}=0$, corresponding
to the vertical axis of the phase diagram. The regions of the phase
diagram traversed by the expanding cooling droplets of QGP formed
in heavy ion collisions with varying energy $\sqrt{s_{NN}}$ are sketched.
The transition from QGP to hadrons is a crossover near the vertical
axis; the thermodynamics of this crossover is well understood from
lattice QCD calculations that are quantitative and controlled in the
yellow region. At higher doping, the transition may become first order
at a critical point. A central goal of the coming second phase of
the RHIC Beam Energy Scan (BES II) is to determine whether such a
critical point exists in the region of the phase diagram that can
be explored using heavy ion collisions. At higher baryon density and
lower temperature, cold dense quark matter is expected to be a color
superconductor. This form of matter may be found at the centers of
neutron stars. (Figs. from \cite{Bazavov:2014pvz,Geesaman:2015fha})\label{fig:lattice}}
\end{figure}

\subsubsection*{1. QCD in Cosmology.}

Heavy ion collisions recreate droplets of the matter that filled the
universe a microsecond or so after the Big Bang. And, it has been
understood since the mid 1970s \cite{Collins:1974ky,Linde:1978px}
that when the universe was only a few microseconds old it was filled
with matter at temperatures above $\Lambda_{{\rm QCD}}$ (the fundamental
energy scale in QCD, of order a few hundred MeV, which is best thought
of as the inverse of the size of a hadron in QCD) and was too hot
for protons, neutrons, or any hadrons to have formed. This direct
and tangible connection to the earliest moments of the universe, together
with the insight that the primordial matter found at these temperatures
had to be some new form of matter not made of hadrons, provides two
powerful motivations for studying ultrarelativistic heavy ion collisions.
Historically, these were the motivations that provided much of the
initial impetus for the field.

In the 1980s, a fair amount of work was done on possible observable
consequences in cosmology of a first order phase transition between
the hot primordial matter and ordinary hadronic matter. These all
relied upon presuming a strong first order phase transition that occurred
via the nucleation of widely separated bubbles of the low temperature
hadronic phase. As the walls of these putative bubbles plowed through
the microseconds-old universe over distances as long as centimeters
or meters, they would have left behind matter that was inhomogeneous
over these long length scales \cite{Witten:1984rs,Applegate:1985qt}.
If this had happened, it would have modified the synthesis of light
nuclei that occurred when the universe was minutes old. Starting in
the late 1990s, and culminating in classic work in the 2000s \cite{Karsch:2001vs,Aoki:2006we},
it became clear from first-principles lattice QCD calculations of
the pressure and energy density of hot QCD matter containing equal
densities of quarks and antiquarks that the transition from primordial
hot QCD matter to hadronic matter in the first few microseconds after
the big bang proceeded via a continuous crossover, not a first order
phase transition. This is in turn consistent with the modern understanding
of big bang nucleosynthesis, which is in accord with cosmological
observations without any of the disruption that a strong first order
phase transition would have introduced \cite{Thomas:1993md}. A continuous
crossover does not introduce any fluctuations on length scales much
longer than the fm-scale natural length scales of QCD, meaning that
it left no imprint in the microseconds-old universe that survived
so as to be visible in some way today. That is, we now understand
that we cannot use cosmological observations to ``see'' the primordial
hot QCD matter that filled the microseconds old universe, or the crossover
transition at which ordinary protons and neutrons first formed.

A central goal of ultrarelativistic heavy ion collisions, then, is
to use these experiments to recreate droplets of Big Bang matter in
the laboratory -{}- where we \textit{can} learn about its material
properties as well as about its phase diagram in ways that we will
never be able to do via observations made with telescopes or satellites.
What can we learn from such studies? What have we learned so far?
One of the most important discoveries made via studying ultrarelativistic
heavy ion collisions is that matter that is a few trillions of degrees
hot is a \textit{liquid}. The early ideas that motivated the field
turned out to be half right: primordial matter at these temperatures
is not made of hadrons, as anticipated; however, at the temperatures
that have been achieved in heavy ion collisions to date, it is not
a weakly coupled plasma of quarks and gluons as originally expected.
Instead, when the hadrons that make up ordinary nuclei are heated
to these extraordinary temperatures, the matter that results is better
thought of as a soup of quarks and gluons, in which there are no hadrons
to be found but in which every quark and gluon is always strongly
coupled to its neighbors, with no quasiparticles that can travel long
distances between discrete scatterings. We shall describe how this
insight was obtained in Section \ref{sec:A-hydrodynamic-fluid}. The
material property that quantifies the liquidness of a liquid made
up of ultrarelativistic constituents is the ratio of its shear viscosity
$\eta$ to its entropy density $s$. \begin{marginnote}[] \entry{Shear viscosity}{The larger the shear viscosity $\eta$ the more easily momentum can be exchanged between distant fluid cells and, consequently, the faster a gradient in fluid velocity (or a sound wave) dissipates into heat.}
\end{marginnote} \begin{marginnote}[] \entry{Specific viscosity}{is the ratio of the shear viscosity to the entropy density, $\eta/s$. It is the natural dimensionless measure of the effects of shear viscosity in a relativistic fluid.}
\end{marginnote}The ratio $\eta/s$ is dimensionless in units in which $\hbar$ and
$k_{B}$ have been set to 1. This ratio plays a central role in the
equations of hydrodynamics where it governs the amount of entropy
produced within the fluid as a sound wave propagates through it, or
more generally as it flows in any nontrivial way. It is the natural
dimensionless measure of the effects of shear viscosity in a relativistic
fluid, and we shall refer to it as the ``specific viscosity''. In
Section \ref{sec:A-hydrodynamic-fluid} we shall sketch how the combination
of data from ultrarelativistic collisions and hydrodynamic calculations
are being used to constrain $\eta/s$, and even its temperature dependence.
We shall also see that $\eta/s$ for the liquid of quarks and gluons
produced in heavy ion collisions is close to the value $1/4\pi$.
Although because of its extraordinarily high temperature this liquid
has extraordinarily large values of both $\eta$ and $s$ relative
to those of any quotidian fluid, its specific viscosity $\eta/s$
is smaller than that of any other known fluid. Interestingly, $1/4\pi$
is the value of the ratio $\eta/s$ in the plasma of infinitely strongly
coupled gauge theories that are cousins of QCD that have a dual gravitational
description in terms of a black hole horizon in 4+1-dimensional Anti-de
Sitter space \cite{Policastro:2001yc,Casalderrey:2014}, a horizon
whose undulations encode the hydrodynamic motion of the plasma \cite{Hubeny:2011hd}.
This connection between the properties of the primordial matter recreated
in heavy ion collisions, via a duality first discovered in string
theory, to properties of black hole horizons certainly provides strong
motivation for pushing the determination of $\eta/s$, and recently
also the bulk viscosity, of quark-gluon plasma to higher accuracy. 

The zeroth order input that a hydrodynamic calculation needs to get
from the microscopic theory of whatever hydrodynamic fluid it is seeking
to describe is the equation of state, relating the pressure of the
fluid to its energy density. In QCD the equation of state, and any
other thermodynamic property of a static volume of quark-gluon plasma
in thermal equilibrium, can be calculated reliably via implementing
the standard path integral formulation of thermodynamics on a discretized
lattice of points in space and Euclidean time, and doing so for a
series of lattices with smaller and smaller spacing between the points,
thus taking the ``continuum limit'' \cite{Lin:2015dga}. Among many
other conclusions, these lattice calculations have taught us that
the transition from the hot, liquid, quark-gluon plasma with zero
baryon number to a gas of hadrons is a crossover \cite{Aoki:2006we},
as in Fig. \ref{fig:lattice}, with no further transitions anticipated
as quark-gluon plasma gradually goes from liquid-like to gas-like
at higher and higher temperatures \cite{Borsanyi:2012ve}. Because
lattice calculations are built upon the Euclidean formulation of equilibrium
thermodynamics, it is much more challenging to use them to gain information
about transport coefficients including the shear and bulk viscosities,
which describe the time-dependent processes via which infinitesimal
perturbations away from equilibrium relax, producing entropy. Pioneering
attempts in these directions have been made \cite{Meyer:2011gj}.
Lattice calculations of more dramatically time-dependent phenomena,
including the quenching of jets in the liquid plasma or the initial
formation of the plasma from a far-from-equilibrium collision, are
beyond the horizon.

The big picture that has emerged over the past 15 years, namely that
the hot matter produced in ultrarelativistic heavy ion collisions
rapidly forms a strongly coupled hydrodynamic liquid with a strikingly
small value of $\eta/s$, has posed new, open, questions that motivate
much experimental and theoretical investigation today. For example,
how, and how quickly, does the hydrodynamic liquid form from the non-hydrodynamic
initial conditions at the moment of the collision? Or, how does a
hydrodynamic liquid emerge at its natural length scales (of order
$1/T$ and longer with $T$ the temperature)\begin{marginnote}[] \entry{Length scale}{According to the uncertainty principle, the characteristic microscopic length scale is proportional to the inverse of the characteristic momentum, which in a relativistic plasma means it is inversely proportional to the temperature.} \end{marginnote}
in an asymptotically free gauge theory in which all matter, when resolved
at short length scales, must be made of weakly coupled quarks and
gluons? Or, what is the smallest droplet of this stuff that can sensibly
be described using the language of hydrodynamics? We will return to
the first and second of these big questions later, in Section \ref{sec:Initial-stage}
for the first and in Section 7 for the second. As to the third big
question, it is currently the subject of intense investigation, both
experimental and theoretical, having been put squarely on the agenda
for the field by measurements made in $pPb$ and $pp$ collisions
at the LHC and $dAu$ collisions at RHIC which indicate that even
proton-sized droplets of hot QCD matter can exhibit liquid-like behavior.
In response to this discovery, theorists have shown that, in the cousins
of QCD with a dual gravitational description, the dynamics of a droplet
of strongly coupled plasma with a temperature $T$ that is $\sim1/T$
in size or larger can be described hydrodynamically \cite{vanderSchee:2012qj,Chesler:2015bba,Chesler:2016ceu}.
This suggests that hydrodynamic behavior should not persist in proton-nucleus
collisions at lower collision energies and hence lower multiplicity.
Noting that this question is at the center of a different article
in this volume \cite{Nagle:2018nvi}, we shall keep our discussion
brief. 

\subsubsection*{2. Phase diagram of QCD.}

Among the most important reasons for studying ultra relativistic collisions
is the expectation that doing so will teach us about the phase diagram
of hot QCD matter, in thermal equilibrium, as a function of both temperature
and baryon doping (see Fig. \ref{fig:lattice}). By baryon doping
(or net baryon number density) we mean the excess of quarks over antiquarks
in the hot matter. The standard parameter used to characterize the
degree of baryon doping is the baryon number chemical potential $\mu_{B}$.
To this point, we have set $\mu_{B}$ to zero, describing matter with
equal densities of quarks and antiquarks. This is a very good approximation
for the matter produced at mid-rapidity in the highest energy heavy
ion collisions at RHIC, an even better approximation at the LHC, and
an exceedingly good approximation in the early universe. In all these
cases, ordinary hadronic matter forms via a continuous crossover as
the liquid QGP expands and cools. However, matter with $\mu_{B}=0$
and varying temperature is only one edge of a phase diagram. A substantial
component of understanding the nature of any complex material in condensed
matter physics is mapping its full phase diagram, and the same is
true in QCD. One way to study QGP doped with a significant excess
of quarks over antiquarks would be to study the debris produced at
very high rapidity in the highest energy heavy ion collisions, the
rapidities where QGP forms from the compressed remnants of the incident
nuclei. Neither RHIC nor the LHC feature detectors that can do this,
at present. Instead, we can scan a region of the phase diagram of
QCD by looking at heavy ion collisions with lower and lower collision
energies in which the initial baryon number found in the incident
nuclei makes a larger and larger contribution to the matter formed
in the collisions: decreasing the collision energy increases $\mu_{B}$,
scanning the phase diagram. Lower energy $AA$ studies are underway
at the SPS \cite{Aduszkiewicz:2017sei} and in the RHIC Beam Energy
Scan \cite{Luo:2015doi}(BES), where tantalizing early results are
in hand from a first phase of the BES program with relatively low
statistics per collision energy \cite{Geesaman:2015fha}. There is
a second high statistics phase of the BES program planned for 2019-2020.
Extensions of this program to even lower collision energies (and hence
even higher $\mu_{B}$, albeit at lower temperature) are planned at
the FAIR facility in Darmstadt, Germany \cite{Heuser:2011zz} and
at the NICA facility in Dubna, Russia \cite{Toneev:2007yu}. One of
the central questions that these experiments aim to answer is whether
the continuous crossover between liquid QGP and hadronic matter turns
into a first order phase transition above some nonzero, critical,
value of $\mu_{B}$, meaning in heavy ion collisions below some collision
energy. There are many models for QCD in which the phase diagram features
a critical point like this \cite{Stephanov:2004wx}. (In QCD with
two massless quarks \textendash{} the ``chiral limit'' \textendash{}
the crossover at $\mu_{B}=0$ becomes a sharp second order phase transition
at which chiral symmetry is restored and a point at $\mu_{B}>0$ where
the transition becomes first order is a tricritical point \cite{Rajagopal:2000wf,Stephanov:2004wx}.)
Furthermore, a critical point has also been seen in some pioneering
efforts to explore physics at nonzero $\mu_{B}$ using lattice techniques,
although because lattice calculations at nonzero $\mu_{B}$ suffer
from a ``sign problem'' these calculations typically require small
$\mu_{B}/T$ and to date it has not been possible to take the continuum
limit \cite{deForcrand:2010ys}. There are also tantalizing indications
of increased non-Gaussian fluctuations \cite{Stephanov:2008qz,Athanasiou:2010kw,Luo:2015ewa}
in exactly the observable that has been predicted to be most sensitive
to critical fluctuations in RHIC collisions near the low end of the
beam energy scan, but these indications are inconclusive given the
presently available statistics. Do we know whether there is a critical
point in the phase diagram at nonzero baryon doping? No. Are there
strong motivations for the experimental program that aims to answer
this question within the next few years? Yes. We have been relatively
brief here, anticipating that the data and analyses coming soon will
warrant a focused review of their own before too long.

\begin{textbox}[t]\section{High baryon density at low temperatures in the cosmos} Pushing to very high baryon doping while staying at low temperature (aka squeezing nuclei without heating them) takes us into another interesting region of the QCD phase diagram. Matter that is sufficiently dense cannot be made of well-separated nucleons, even at low temperatures: the nucleons are crushed into one another. Because quarks attract each other, cold, dense matter in which quarks fill momentum space up to some high Fermi momentum is a color superconductor in which a condensate of correlated Cooper pairs of quarks creates a superfluid and yields the QCD-analogue of a Meissner effect. Extensive theoretical analyses of the phase diagram and consequent properties of color superconducting quark matter \cite{Alford:2007xm} have been performed; they are well understood at asymptotic densities, but at densities of order 10 times that of nuclei they turn out to be sensitive to the ratio of the strange quark mass to the Fermi momentum as well as to the strength of the Cooper pairing, making them hard to pin down quantitatively. Experimental data is sorely needed. Unfortunately, the only place in the universe where cold dense quark matter may be found is in the centers of neutron stars. Remarkably, the first collision between two neutron stars has just been observed by the LIGO and VIRGO collaborations, via the gravitational waves it produced\cite{TheLIGOScientific:2017qsa}! Although the gravitational waves from this discovery event seen by LIGO only reveal the inspiraling incident neutron stars, with coming improvements to LIGO's sensitivity future events will give us a view of the collision itself, making it possible to learn about the compactness and density profile of the incident neutron stars and, conceivably, whether or not they feature dense quark matter cores. If they do, present constraints on heat transport in neutron stars coming from X-ray observations of how they cool will turn into constraints on the transport properties of cold dense quark matter. \end{textbox} 

\subsubsection*{3. Emergence of complex quantum matter.}

In the history of the universe, liquid quark-gluon plasma was the
earliest complex form of matter to form. At much earlier times, when
the temperature was a few orders of magnitude hotter than those of
interest to us here, the matter that filled the universe \textit{was}
a weakly coupled plasma of quarks and gluons. We know this because
QCD is asymptotically free,\begin{marginnote}[] \entry{Asymptotic freedom} {means that the QCD coupling $\alpha_s$  weakens for interactions between quarks that are close together (or scatter at high energy), and is strong and nonperturbative at length (or energy) scales of order the (inverse) of the size of a hadron. In gas-like QGP at asymptotically high temperatures, the interaction energy is small compared to the kinetic energy.}\end{marginnote}
meaning that quarks and gluons interact with each other only weakly
when they scatter off each other with large enough momentum transfer.
Not only was liquid QGP the earliest complex matter to form, there
is also a sense in which it is the simplest form of complex matter
that we know of, namely the complex matter that is ``closest'',
most directly connected to, the fundamental laws that govern all matter
in the universe, in this case the fundamental theory of QCD. Again
because QCD is asymptotically free, we know that if we could hold
a droplet of the liquid QGP with temperature $T$ in place and study
its microscopic structure with a spatial resolution that is much finer
than $1/T$, for example via scattering high energy electrons off
it in this thought experiment, what we would see is weakly coupled
quarks and gluons. This is the genesis of the strongest motivation
for developing experimental techniques for probing the structure of
the liquid QGP on varying length scales. We know that at the shortest
length scales we must see weakly coupled quarks and gluons.We also
know that at length scales of order $1/T$ and longer we see a liquid
in which neighboring ``unit cells'' are tightly coupled to each
other, meaning that the liquid flows hydrodynamically with a small
$\eta/s$. If we can probe both these length scales and scales in
between, for example via studying how jets, (which are intrinsically
multiscale probes), or heavy quarks with varying mass, or tightly
bound quarkonium mesons with varying sizes, ``see'' the plasma and
how the plasma responds to their passage through it, we have a chance
to probe, and maybe even \textit{understand}, how the simplest form
of complex matter that we know emerges from weakly coupled, asymptotically
free, constituents at short length scales. The question of how the
almost infinite variety of complex forms of matter that we see in
the world around us emerge from laws of nature that are so simple
that they can easily fit on a T-shirt is one of the great quests of
modern physics. If we can answer it for the case of liquid quark-gluon
plasma, which we have a chance to do by virtue of this simple form
of complex matter being so close to its laws-of-nature underpinnings,
maybe we have a chance of shedding light on the larger more general
question.

\section{Phenomenology of heavy ion collisions\label{sec:Phenomenology-of-heavy}}

In the study of heavy ion collisions, experimenters have only two
quantities under their direct control: which two nuclei they collide
and at what energies. The energies are known to high precision. However,
knowing the colliding nuclei is not the same as knowing the colliding
systems. Neither the impact parameter $b$ (the transverse distance
between the center of masses of the two nuclei) nor the location and
motion of the nucleons in the nuclei, let alone that of the quarks
and gluons in the nucleons, are measurable quantities. They have to
be inferred, as best as possible or as needed, event by event, from
the observed outcome of the collision. This then makes it possible,
after the fact, to select an ensemble of collisions with a relatively
narrow distribution of impact parameters.

Based on nuclear and particle physics studies we know that, from the
point of view of relativistic heavy ion collision studies, the nuclei
can instantaneously be reasonably well approximated by a collection
of nucleons, distributed on average according to a well-determined
three-dimensional distribution. We also know the average quark and
gluon content of the nucleons in the nuclei in terms of parton distribution
functions or PDFs, and find that the PDFs in nuclei differ only mildly
from those describing free nucleons \cite{Eskola:2016oht}. Furthermore
we can use the measured energy dependent total inelastic $pp$ cross-sections
$\sigma_{pp}(\sqrt{s})$ \cite{Patrignani:2016xqp} to model the nucleons
in the nucleus as hard spheres with a radius that depends on energy.

\begin{figure}
\centering{}\includegraphics[width=1.2\columnwidth]{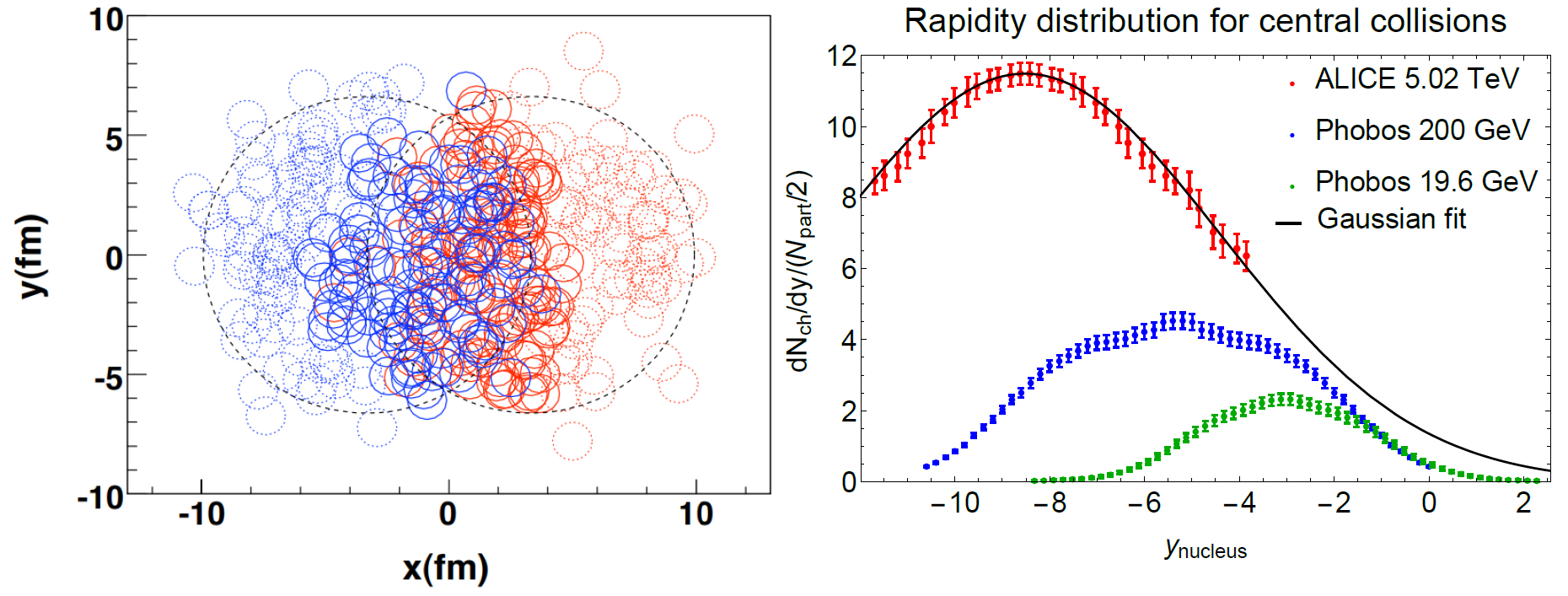}\caption{(left) An example of a PbPb collision at LHC with impact parameter
$b\approx7$ fm. Number of participants (solid) are counted by nucleons
that collide with any nucleon, whereas the number of binary collisions
count all overlapping blue/red nucleon pairs. Spectators (dashed)
do not collide. (Fig. from \cite{Alver:2008aq}.) (right) Rapidity
distributions of charged hadrons, in the rest frame of one of the
nuclei, for AuAu collisions at 19.6 and 200 GeV (converted from pseudorapidity
from \cite{Alver:2010ck} using a simplified Jacobian) and for PbPb
collisions at 5.02 TeV \cite{Adam:2016ddh}.\label{fig:glauber-and-rapidity}}
\end{figure}

It will turn out to be useful to do a ``gedanken experiment'' where
we imagine the colliding nuclei to be composed of $A$ (transparent)
spheres of radius $\sqrt{\sigma_{pp}/4\pi}$, where $A_{L,R}$ is
the number of nucleons inside the left- and right-moving nuclei. We
then call those nucleons that do not encounter any nucleon from the
other nucleus spectators (dashed in Fig. \ref{fig:glauber-and-rapidity}).
These nucleons continue traveling down the beam pipe, and the number
of spectators $N_{\text{spec}}$ can hence in principle be measured
directly, although in practice this is usually hard. In the gedanken
experiment, all other `wounded' or participating nucleons collide
with at least one other nucleon and make up the number $N_{\text{part}}$
(solid in Fig. \ref{fig:glauber-and-rapidity}, by definition $N_{\text{spec}}+N_{\text{part}}=A_{L}+A_{R})$.
It is unfortunate that $N_{\text{spec}}$ is not measurable in practice,
since if it were then $N_{\text{part}}$ could be determined directly.
Lastly, if we imagine the spheres as transparent we can also count
the total number of encounters between left- and right-moving nucleons,
which we will call the number of binary collisions $N_{\text{coll}}$.
For example, if one \textquotedbl{}nucleus\textquotedbl{} consists
of 7 nucleons lined up in a row and it collides head-on with a ``nucleus''
consisting of 4 nucleons in a row, $N_{\text{part}}$ = 11, $N_{\text{coll}}$
= 28, $N_{\text{spec}}$ = 0 and the impact parameter $b=0$. In a
real central heavy ion collision a nucleon at the center of one nucleus
will on average hit about $12$ nucleons from the other, but less
if it is located at the edge of the collision. So, $N_{\text{coll}}$
will then be much larger than $N_{\text{part}}$, and even more so
for the more central collisions.\begin{marginnote}[] \entry{Participant}{Nucleon that collides with at least one other nucleon.}\entry{Spectator}{Nucleon that does not collide and hence keeps on moving along the beam direction.} \entry{Binary collisions}{Total number of nucleon pairs that collide, assuming transparency of the collision.}\end{marginnote}

In a $pA$ collision, the probability of the proton hitting another
nucleon is given by the ratio $\sigma_{pp}/\sigma_{pA}$ of inelastic
scattering cross-sections. This makes it possible to determine that
on average $N_{\text{coll}}=N_{\text{part}}-1=A\,\sigma_{pp}/\sigma_{pA}$,
which can be measured directly. Experimental data on $pA$ collisions
with widely varying $A$ and collision energies (going back to the
1970s \cite{Busza:1975te}) show that the number of particles produced
in such collisions is proportional to $N_{\text{part}}$ to a good
approximation. Although in $AA$ collisions $N_{\text{part}}$ and
$N_{\text{coll}}$ cannot be determined directly from measured cross-sections,
there is a well-defined theoretical procedure \cite{Miller:2007ri}
(called a ``Glauber Model Calculation'') for determining these abstract
measures, at least on average within ``centrality classes''. This
procedure generates many configurations with different $b$, as illustrated
in Fig. \ref{fig:glauber-and-rapidity}, and thereby generates Monte
Carlo distributions of $N_{\text{part}}$ (as well as $N_{\text{coll}}$).
It is then assumed that there is a monotonic relation between the
number (or energy) of the produced particles and $N_{\text{part}}$.
For example, it is assumed that events, in which the number (or energy)
of particles falls into the highest 5\% class, correspond to the 5\%
most central collisions, with $N_{\text{part}}$ or $N_{\text{coll}}$
(from the Glauber Model Calculation) in the highest 5\% category.
The bases for this prescription are, first, the experience from $pA$
collisions that we mentioned above and, second, the observation that
the shape of the measured probability distribution for the number
(or energy) of particles in an ensemble of $AA$ events is similar
to the probability distribution for $N_{\text{part}}$ obtained from
a Glauber Model calculation. Most important, the participant scaling
observed for collisions of nuclei with widely varying $A$ (discussed
below) provides a strong indication that these abstract measures in
some way reflect a physical reality.

We now describe, as a function of energy, $N_{\text{part}}$ and $N_{\text{coll}}$,
the most general features observed when two heavy ions collide at
relativistic velocities. In order to give the ``big picture'', in
this discussion we shall ignore small differences and subtle effects.
For useful summaries of and references to RHIC and LHC data we refer
the reader to \cite{Arsene:2004fa,Back:2004je,Adams:2005dq,Adcox:2004mh}
and \cite{Armesto:2015ioy,Foka:2016vta,Foka:2016zdb}. See also recent
proceedings of Quark Matter conferences and \cite{Geesaman:2015fha,LongRangeEU}
for an overview of theoretical and experimental work.

\subsubsection*{Hard collisions}

High-$p_{T}$ $\gamma$ and $Z^{0}$ production have been studied
in $pp$ and $AA$ collisions \cite{Aad:2012ew,Chatrchyan:2014csa,Aad:2015lcb}.
The measured $pp$ cross-sections are well understood. They are in
excellent agreement with predictions based on the known PDF's and
perturbative QCD theory (pQCD). The measured $AA$ production rates,
in turn, are in excellent agreement with the product of $N_{\text{coll}}$
and the yield in a single $pp$ collision, taking into account the
measured modifications of the PDF's of nucleons inside nuclei and
uncertainties in the determination of $N_{\text{coll}}$. Since the
hard gammas and $Z^{0}$'s are not affected by the post-collision
$AA$ environment (both are colorless), these results show that we
have a good understanding of the initial hard (large $p_{T}$) parton-parton
interactions in $AA$ collisions and of the determination of $N_{\text{coll}}$,
meaning that these results provide independent confirmation from experimental
data of the Glauber Model Calculations described above. This, in turn,
implies that the $AA-pp$ comparisons of outgoing strongly interacting
(colored) hard probes (jets, high-$p_{T}$ hadrons, heavy quarks,
etc) can be used to give us valuable information about the nature
of the medium produced in heavy ion collisions and on how the colored
hard probes themselves are modified as they traverse the medium produced
in $AA$ collisions. Experimental measurements of jets themselves,
as well as of high-$p_{T}$ hadrons and heavy quarks which come from
jets \cite{Connors:2017ptx}, all show that jets lose considerable
energy as they propagate through QGP, with the ``lost'' energy ending
up as many soft particles ($\lesssim3$ GeV/c \cite{CMSsoft}) moving
at large angles relative to the original jet direction \cite{Khachatryan:2015lha},
suggesting that the jet leaves a wake behind in the liquid QGP. This
suite of results and phenomena, collectively referred to as ``jet
quenching'', are very important since they give us direct evidence
of very strong interactions occurring after the collision, strong
interactions between the jet and the liquid as well as strong interactions
within the perturbed liquid; we shall return to them in Section \ref{sec:Jets-in-quark-gluon}. 

\subsubsection*{Baryon stopping power:}

It has long been known that in lower energy $pp$ collisions the longitudinal
momentum of forward going protons in the final state have a flat distribution,
evenly distributed between 0 and the incident energy \cite{Brenner:1981kf,Barton:1982dg}.
This implies that, on average, a proton loses half its energy, which
is about one unit of rapidity. In $pA$ and $AA$ collisions, the
energy lost by the incident nucleons is higher on average, and more
narrowly distributed. On average, in $AA$ collisions, each participant
loses about two units of rapidity \cite{Busza:1989px}, which is to
say 85\% of its energy goes into the creation and kinetic energy of
a very large number of particles, up to 30,000 in central PbPb collisions
at the LHC. The net proton rapidity distribution in $AA$ collisions
has a double hump structure \cite{Arsene:2009aa}, each consisting
of hot baryonic matter moving at a speed of about two units of rapidity
below that of the incident beam, and having a net baryon density of
about 5-10 times that of normal nuclear matter \cite{Busza:1983rj}.
(At LHC the beam rapidity is at $y=8.5$. As seen in the frame in
which $y=6.5$ is at rest, an incident disc that is Lorentz contracted
by a factor of about $\cosh(2)$ is hit by a disc that is Lorentz
contracted by about a factor of $\cosh(15)$ and brought approximately
to rest, compressed by roughly $2\cosh(2)\approx7.5$.) 

A further consequence is that the maximum value of the net baryon
density at mid-rapidity is produced when heavy ions collide with $\sqrt{s_{NN}}\approx7$
GeV. Above this collision energy, the mid-rapidity net baryon density,
and so also the baryon chemical potential in the QGP produced at mid-rapidity,
decreases with energy. By top RHIC collision energies, and even more
so for LHC energies, both are essentially zero \cite{Abelev:2013vea}.

\subsubsection*{Energy and centrality dependence of multi-particle production:}

For practical reasons we have most information about charged particles,
which corresponds to about 2/3 of all the produced particles. However
there is no reason to doubt that the picture obtained from the charged
particles is anything other than the whole picture!

From the lowest energies measured \cite{Elias:1979cp}, through RHIC
\cite{Alver:2010ck} to LHC \cite{Adam:2016ddh} energies, for all
$AA$, $pA$, $\pi A$ and $KA$ collisions, the total number of charged
particles produced is approximately proportional to the number of
participants. This is known as ``participant scaling'' and is not
well understood. Even more surprisingly, provided that one takes into
account the fraction of energy that is taken away by the forward going
baryons and not available for particle production (mentioned above
and see \cite{Brenner:1981kf}), the total number of produced charged
particles per participant in $AA$ collisions is the same as that
in $pp$ and $e^{+}e^{-}$ collisions \cite{Back:2006yw}. However
this arises, it suggests that on average in $AA$ collisions most
of the entropy production, which is proportional to the number of
produced particles\footnote{At the chemical freeze-out temperature, the multiplicity of each of
the hadron species present in QCD is given to a good approximation
by a thermal distribution. (We shall discuss this further below and
in Section \ref{sec:A-hydrodynamic-fluid}.) This makes a direct connection
between the entropy at this moment and the number of charged particles
possible. The contribution from any single species in a thermal distribution
to $N_{ch}$ is proportional to $S$, with a proportionality constant
that decreases with increasing mass. Adding up all the known species
of hadrons yields $N_{\text{ch}}\approx S/7.25$ at freeze-out \cite{Muller:2005en}.} and hence the number of participants, occurs early in the collision
and that there is little, if any, late stage entropy production. We
shall later see that there are powerful arguments in support of this
conclusion: after the early stage of the collision when entropy production
is copious, a hydrodynamic fluid forms, and because this fluid has
low specific viscosity, little entropy is produced subsequently, as
the liquid flows.

For a given number of participants, the total number of produced particles
$N$ increases with energy (as $N\propto s_{NN}^{0.15}\log(s_{NN})$)\cite{Abbas:2013bpa}.
Except for close to the receding discs, the longitudinal rapidity
distributions look approximately like wide Gaussians with a width
which increases as $\log(s_{NN})$ (i.e. as the beam rapidity or longitudinal
phase space, see Fig. \ref{fig:glauber-and-rapidity}) \cite{Adam:2016ddh},
and increases weakly from central to peripheral collisions \cite{Back:2002wb}.
The produced particle density $dN/dy$ has hence no boost invariant
plateau and the maximum at mid-rapidity increases with energy as ($dN/dy\propto s_{NN}^{0.15}$)\cite{Aamodt:2010pb}.
The simplicity of these empirical facts at first glance seems to be
at odds with the complex sequence of stages that precedes particle
production. In the rest frame of the produced particles, the finally
observed particle density $dN/dy$ is the result of a local history
which includes the initial impact of the nuclei, followed by the creation,
expansion and flow of a hot medium, and its eventual hadronization
into particles. At one level, the simplicity of the empirical facts
can be explained by noting that the number of particles in the final
state is proportional to its entropy and concluding that at any rapidity
most of the entropy is produced very early in the collision, making
$dN/dy$ insensitive to all that happens later. However, these facts
are nevertheless not fully understood. For example, the energy dependence
and centrality dependence are surprisingly independent of each other
from the lowest to the highest energies studied \cite{Alver:2010ck,ATLAS:2011ag,Adam:2015ptt}.
This means that the naively expected increase with energy resulting
from the increase of hard (proportional to $N_{\text{coll}}$) relative
to soft (proportional to $N_{\text{part}}$) processes does not play
a leading role in determining the number of produced particles. 

Another interesting observation is the so-called extended longitudinal
scaling. If the rapidity of one nucleus is kept constant and that
of the other is gradually increased, we see that at first $dN/dy$
increases, but it then reaches a limiting value (see Fig. \ref{fig:glauber-and-rapidity}).
Thinking of the second nucleus as a wall of gluons, boosting these
gluons more and more seems to have no effect on particle production
in the collision around the rapidity of the first nucleus. This phenomenon
has been observed for all systems studied \cite{Busza:2011pc} and
is direct evidence that a kind of saturation occurs in the fast nucleus
\cite{Gelis:2006tb}.

Finally, we point out that all these facts do not support Landau's
\cite{Landau:1953gs,Florkowski:2010zz} and Fermi's \cite{Fermi:1950jd}
early models, in which they postulated that the two colliding systems
completely stop each other and then (in the case of Landau after a
period of hydrodynamic expansion from rest) break up into particles
according to thermodynamic laws. And, they are also inconsistent with
Feynman's intuition \cite{Feynman:1969ej} that $dN/dy$ at mid-rapidity
would not increase with increasing collision energy. Feynman expected
the rapidity-distribution of the produced particles to broaden with
increasing collision energy; this does happen, but, because of the
rapid rise of the gluon PDF which Feynman did not anticipate, the
total particle production increases fast enough that $dN/dy$ at mid-rapidity
nevertheless increases.

\begin{figure}
\centering{}\includegraphics[width=1.2\textwidth]{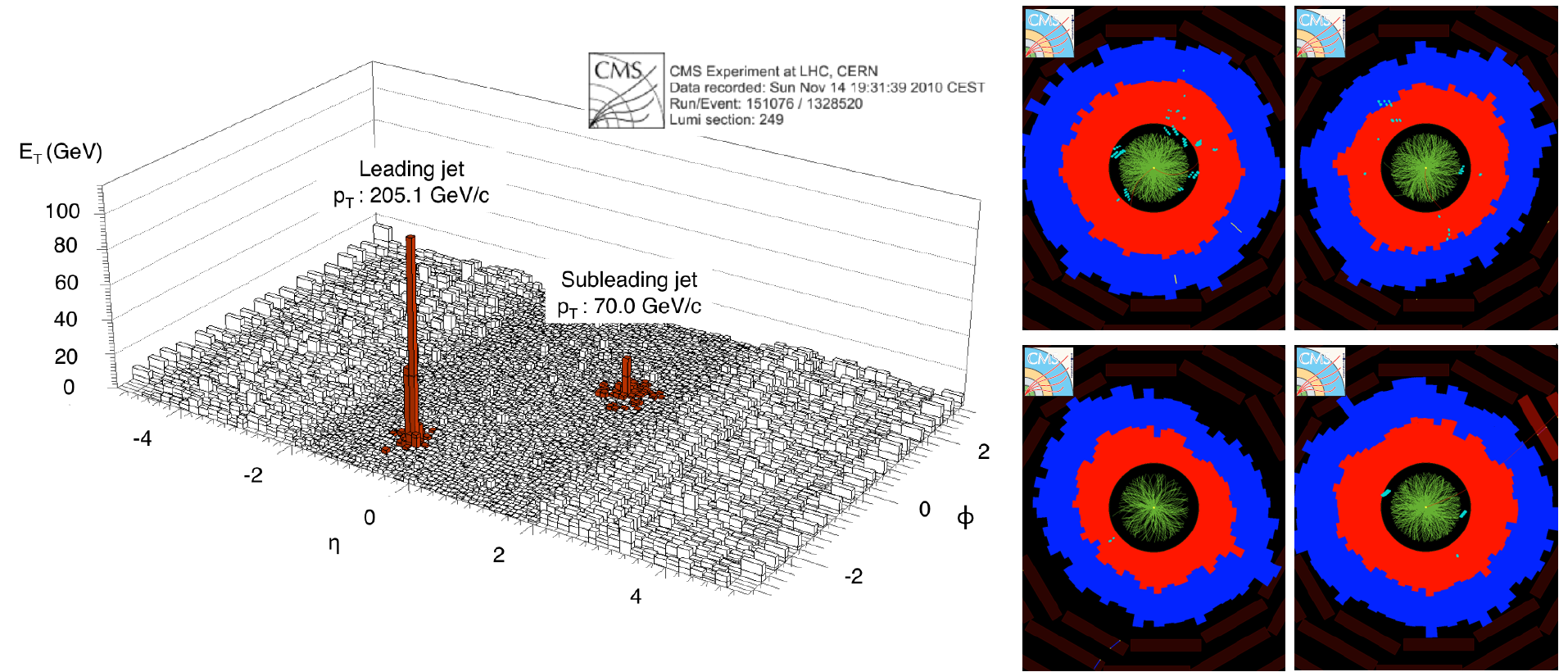}\caption{(left) This event display \cite{Chatrchyan:2011sx} shows energy deposited
in the CMS calorimeter in a heavy ion collision as a function of azimuthal
angle $\phi$ and pseudorapidity $\eta$, a proxy for rapidity which
is more easily measured. Two jets of very different energies are apparent,
suggesting that one jet lost more energy as it traversed the droplet
of QGP. (right) CMS event displays showing azimuthal distribution
of charged tracks (green) and energy in the electromagnetic and hadronic
calorimeters (red and blue respectively) from four heavy ion collision
events as seen by the CMS detector. The azimuthal anisotropies are
apparent, with the upper-right and lower-left events showing marked
ellipticity and the bottom-right event showing a substantial anisotropy
in a higher harmonic. It is remarkable that the strongly coupled character
(left) and the liquid nature (right) of the QGP formed in these collisions
can be seen so clearly in individual events. \label{fig:dijet}}
\end{figure}

\subsubsection*{Particle correlations:}

Strong correlations are observed between particles produced with momenta
in different directions. They are much stronger than expected from
the superposition of independent $pp$ collisions, and are evidence
that the products of the initial collision act collectively. 

Correlations between particles that are widely separated in rapidity
are observed \cite{Chatrchyan:2012wg,Aamodt:2011by}. By causality,
they must have their origin in early times and thus give information
about correlations present at the earliest stages of the collision
of the two nuclei. Azimuthal correlations as in Fig. \ref{fig:dijet}
(right), in particular, have a very pronounced and rich structure
and have been extensively studied as a function of the centrality
of the collision, produced particle type, rapidity, transverse momentum,
and expected event-by-event geometrical fluctuations of the nuclei
\cite{Heinz:2013th,Li:2017qvf}. As discussed in detail in Section
\ref{sec:A-hydrodynamic-fluid} they can be remarkably well explained
by relativistic hydrodynamics, if one assumes that in high energy
heavy ion collisions, before the final production of free streaming
particles, some kind of a relativistic liquid is formed, which expands,
flows radially at about half the speed of light, and in which pressure-driven
anisotropies in the flow velocity form and persist because the liquid
has an incredibly low viscosity to entropy ratio, in fact lower than
that of any other known liquid. It is for this reason that we know
that QGP is a strongly coupled liquid. 

A good way to see that the medium produced in a heavy ion collision
indeed behaves like a low viscosity hydrodynamic liquid is to note
the following. Like all nuclei, those that collide in heavy ion collisions
are lumpy, meaning that the energy density of the matter produced
in the earliest moments of the collision must also be lumpy. If that
matter were a tenuous gas-like plasma, made of lots of particles that
fly around while interacting only rarely with each other, the initial
lumpiness would quickly disappear as the particles fly around in random
directions, and at the end of the day all you would see is an isotropic
explosion of particles, with just as many particles going in any direction
as in any other. If, instead, the matter that is produced is a liquid
whose motion is governed by hydrodynamics, the initial lumpiness will
mean initial pressure gradients, and these pressure gradients will
drive anisotropic flow in the liquid. If the viscosity of the liquid
is large, these anisotropic flows will damp out. Instead, what is
seen in heavy ion collisions is substantial anisotropies in the azimuthal
distribution of particles in the final state (as in Fig. \ref{fig:dijet}),
which reflect azimuthal anisotropies in the geometry of the overlap
region of the colliding nuclei. This means that the matter produced
in the collisions must be a fluid with low specific viscosity.

\subsubsection*{Medium properties:}

Azimuthal correlations give information about the relativistic hydrodynamic
nature of the medium, its transport coefficients, and about the fluctuations
in the initial state from which it forms \cite{Heinz:2013th,Romatschke:2017ejr,Braun-Munzinger:2015hba}.
Jet-quenching studies \cite{Mehtar-Tani:2013pia,Qin:2015srf,Connors:2017ptx}
give us a wealth of information on both how the medium responds when
a high energy quark or gluon jet produced in an initial hard scattering
traverses it, and how a fast quark or gluon jet are modified by the
medium as they pass through it. As mentioned earlier, it is jet quenching
that shows that QGP is extremely strongly interacting and is giving
us much insight into the workings of QCD. So we return to this important
topic later, in Section \ref{sec:Jets-in-quark-gluon}. There are
no known measurements that give us direct and unambiguous information
about what is the nature of the produced low viscosity fluid, is it
in equilibrium, how does it form and how does it equilibrate, what
is its equation of state and phase diagram, what are the best degrees
of freedom for its description, how many thermodynamic degrees of
freedom does it have compared to a hadron gas, and is it a liquid
of deconfined quarks and gluons. However there are indirect measurements
that give us insight into these questions and, together with theoretical
studies, particularly lattice gauge calculations, a consistent picture
of the nature of QGP is emerging.

For example, near mid-rapidity, in RHIC or LHC central $AA$ collisions,
the ratios of the hadrons containing the lighter $u$, $d$, and even
the $s$ valence quarks \cite{Koch:2017pda} are well represented
by a system in chemical equilibrium at a temperature of about 155
MeV \cite{Becattini:2012xb,Andronic:2017pug}. (Note that the number
densities of charm and bottom quarks do not reach chemical equilibrium
because the temperature is not high enough, meaning that their multiplicities
retain memory of their initial production. Top quarks are not relevant
here because of their short lifetimes.) On the other hand, the transverse
momentum spectra are consistent with a system in equilibrium with
a temperature of about 95 MeV and substantial radial flow \cite{Abelev:2013vea,Adamczyk:2017iwn}.
Consistent with QGP being a strongly coupled liquid which behaves
hydrodynamically as it expands and cools, there are no indications
of any abnormal production of very low momentum pions (wavelength
\ensuremath{\sim} size of QGP droplet) \cite{Back:2004zx}, for example
from the formation of a region of disoriented chiral condensate \cite{Rajagopal:1995bc}.

These facts, combined with the observed azimuthal correlations, participant
scaling and jet quenching, are consistent with the following interpretation.
Very early in the collision of the two Lorentz contracted nuclei,
a thin cylindrical volume of QGP liquid is formed, with an entropy
that is determined early, before the fluid hydrodynamizes. At first
this liquid has a non-uniform energy density and temperature distribution
determined by the lumpiness of the colliding nuclei. It expands and
cools in accordance with relativistic hydrodynamics, and because its
specific viscosity is so small it does so almost isentropically. When
the temperature of the system locally falls below about 155 MeV, the
QGP goes through a crossover phase transition and hadronizes. It is
not known whether the hadrons are produced in chemical equilibrium
or chemically equilibrate quickly, after the phase transition. All
this is the so-called chemical freeze-out. The produced hadronic system
then continues to interact, expand and cool until the temperature
falls to about 95 MeV when thermal freeze out occurs. After thermal
freeze out, the hadrons stream outwards freely, eventually reaching
the detectors. At the thermal freeze out time, in addition to thermal
motion the hadrons have radial and anisotropic velocities inherited
from the flow of the expanding liquid that came before.

\begin{figure}
\centering{}\includegraphics[width=6.8cm]{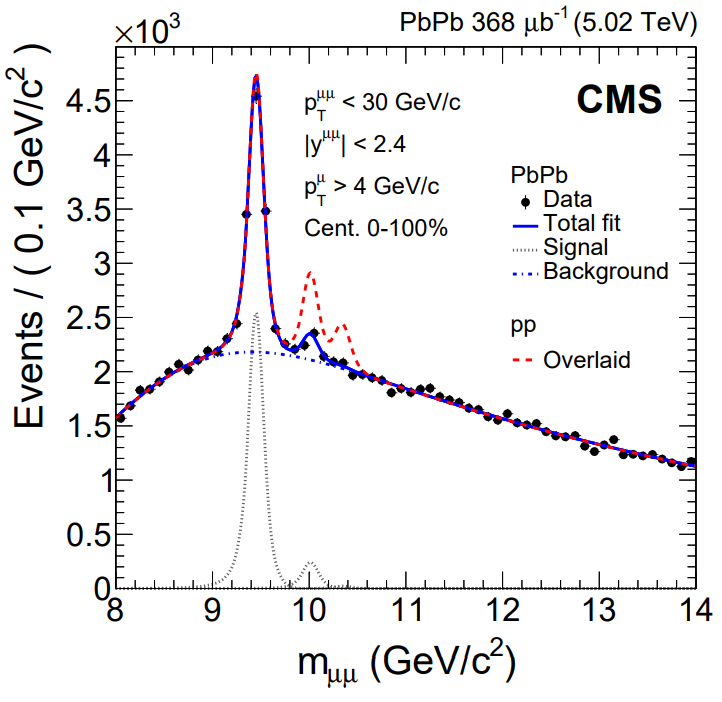}\caption{The dimuon invariant mass distribution shows the different $\Upsilon$
states, whereby the red dashed line shows the $pp$ result added to
the PbPb background and normalized to the $\Upsilon$(1S) state \cite{Sirunyan:2017lzi}.
Clearly the $\Upsilon$(2S) and $\Upsilon$(3S)\emph{ }states in PbPb
collisions are much less pronounced, which is interpreted as the melting
of these larger and less strongly bound $b\bar{b}$ states when they
find themselves immersed in QGP.\label{fig:upsilons}}
\end{figure}
Measurements of quarkonia ($J/\psi$ and $\Upsilon$ mesons made from
moderately heavy 1.3 GeV charm and heavy 4.2 GeV bottom quarks respectively)
production in heavy ion collisions compared to that in $pp$ collisions
\cite{Foka:2016zdb} provide further information about the properties
of the QGP medium, in two different ways. Consider first the case
in which the production of a heavy $Q\bar{Q}$ pair in the hard collisions
at the very beginning of the collision process is rare, for example
as for $b\bar{b}$ pairs in LHC collisions, ideally meaning that in
each heavy ion collision there are zero or one $b\bar{b}$ pairs.
The $b\bar{b}$ pair finds itself immersed in the QGP medium which,
via Debye screening, weakens the attractive force between the pair.
The smallest, most tightly bound, $\Upsilon$(1S) state has a size
comparable to or even smaller than the Debye length of QGP, meaning
that the $b$ and $\bar{b}$ may be close enough together to remain
bound even when immersed in QGP. The $\Upsilon$(3S), on the other
hand, is comparable in size to ordinary hadrons meaning that a $b$
and $\bar{b}$ with this separation do not attract each other when
screened by QGP, and drift apart. The $\Upsilon$(2S) is an intermediate
case. Figure \ref{fig:upsilons} is a beautiful example of data which
shows that $\Upsilon$ states with different sizes and binding strengths
do indeed have different probabilities of surviving in QGP, supporting
this picture.

$J/\psi$ production in LHC heavy ion collisions is interestingly
different \cite{Foka:2016zdb}. These collisions are sufficiently
energetic that, on average, about 30 $c\bar{c}$ pairs are produced
in each heavy ion collision \cite{Zhang:2007dm}. In $N_{{\rm coll}}$
independent $pp$ collisions, in which the same number of $c\bar{c}$
pairs are produced, any $J/\psi$'s that form originate from the $c$
and $\bar{c}$ produced in a single hard scattering. In a heavy ion
collision, those primordial $J/\psi$ are expected to fall apart in
QGP, as above. However, it now becomes possible for a $J/\psi$ to
form via a new process in which a $c$ and $\bar{c}$ from different
initial hard scatterings thermalize in, and diffuse through, the QGP
formed in the collision and then happen to find each other at the
time of hadronization. $c\bar{c}$ production is so copious at LHC
energies that there are more $J/\psi$'s produced in heavy ion collisions
via this recombination process than are produced in the standard fashion
in $N_{{\rm coll}}$ independent $pp$ collisions. This confirms that
the $c$ and $\bar{c}$ quarks produced in heavy ion collisions wander
independently of each other, and is thus a direct confirmation that
quarks in QGP are not confined within hadrons.

\subsubsection*{Comparing $AA$ collisions with $pp$ and $pA$:}

Unlike in $AA$ collisions, the jet quenching phenomenon is not seen
in $pA$ collisions: at mid-rapidity the number of jets seen is just
what one would expect from $N_{{\rm coll}}$ $pp$ collisions \cite{Khachatryan:2016xdg}.
(At large forward and backward rapidities there are deviations from
this, deviations that are understood as coming from differences between
nuclear and nucleon initial states \cite{Khachatryan:2016xdg}.) This
absence of jet quenching came as no surprise since $pA$ collisions
produce an energetic final state that is small in transverse extent
and because in ultrarelativistic collisions the incident nucleus is
highly contracted in the longitudinal direction the nascent jets are
quickly outside the energetic final state and cannot encounter the
spectators from the incident nucleus. What did come as a surprise
is how many other phenomena are similar in $AA$ and $pA$ collisions,
and even in $pp$ collisions, in particular when the comparison is
done between collisions in different systems with the same final state
particle density $dN/dy$. Examples include the rapidity distribution
\cite{Busza:2004mc}, particle spectra \cite{Adam:2016dau}, particle
ratios including those involving strangeness \cite{ALICE:2017jyt},
and, most significantly, the azimuthal anisotropies \cite{Bozek:2013uha,Aaboud:2016yar}
encoded in multiparticle correlations that were once thought to be
unique to $AA$ collisions. Although these similarities are not yet
well understood and are currently topics of intense debate, it is
tempting to interpret them as indicating that proton-sized droplets
of QGP can be formed in those $pp$ and $pA$ collisions that produce
final states with sufficiently large $dN/dy$. This has prompted a
recent theoretical focus on the question of how small the smallest
droplet of QGP that can be described hydrodynamically can be, and
the realization that in the case of a strongly coupled liquid-like
QGP the answer seems to be around $1/T$ \cite{Chesler:2015bba,Chesler:2016ceu,vanderSchee:2012qj}.
This makes it plausible after the fact that some sufficiently energetic
$pp$ or $pA$ collisions can make droplets of QGP with temperatures
well in excess of the inverse of the proton size. A full discussion
of the ``heavy ion'' features seen in small collision systems can
be found in a companion article in this volume \cite{Nagle:2018nvi}.

\section{A hydrodynamic fluid\label{sec:A-hydrodynamic-fluid}}

\begin{figure}
\centering{}\includegraphics[width=1.2\textwidth]{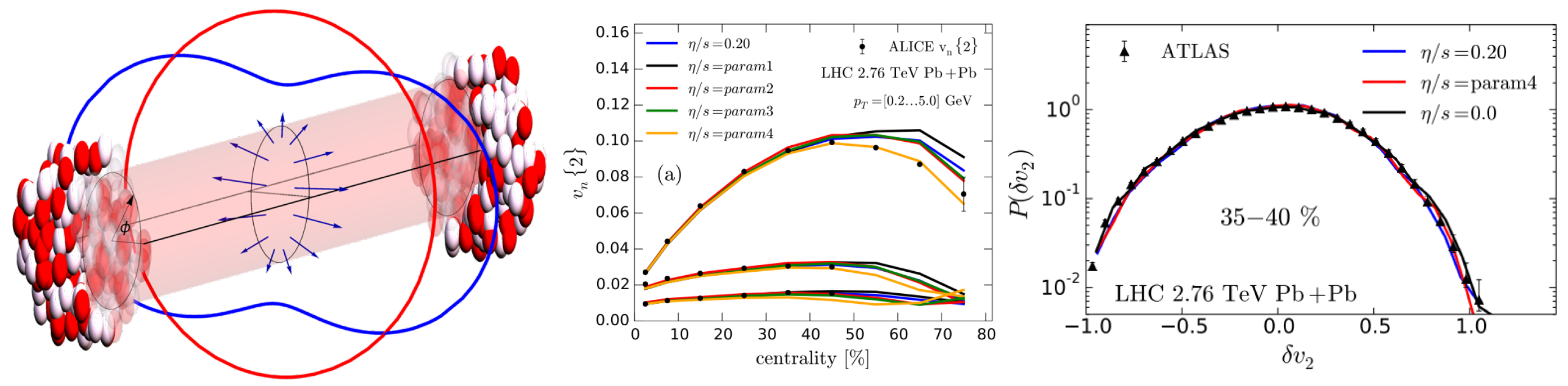}\caption{(left) A peripheral heavy ion collision produces an approximately
elliptical collision region (shaded red). A gas of weakly interacting
particles would give a more or less isotropic distribution of final
particles (red), whereas a fluid would give rise to an anisotropic
distribution (blue), due to the difference in pressure gradients in
the transverse directions. (middle) In \cite{Niemi:2015qia} a hydrodynamic
model with several temperature-dependent parametrizations of $\eta/s$
(see paper) is compared with ALICE measurements of the anisotropy
\cite{ALICE:2011ab}, as obtained by the integrated Fourier coefficients
$v_{n}$ ($n=$2 to 4 from top to bottom), for $\sqrt{s_{NN}}=2.76$
TeV collisions as a function of the centrality class (0\% being head-on
collisions). For more off-central collisions there is an increasing
and large $v_{2}$, giving a hint into the importance of hydrodynamic
evolution. (right) We show event-by-event distributions of the $v_{2}$
distribution for off-central collisions from \cite{Niemi:2015qia}
compared to ATLAS measurements \cite{Aad:2013xma}. In this Section,
we shall discuss the comparison between precise measurements of the
anisotropy and increasingly sophisticated hydrodynamic calculations,
as in the middle and right figures.\emph{ \label{fig:elliptic-flow}}}
\end{figure}

A crucial feature in our description of a heavy ion collision and
the interpretation of the observed facts is that shortly after the
initial impact of the heavy ions and before the hadronization process,
the system (QGP) is in the form of a near perfect (extremely low specific
viscosity) liquid. We now address in more detail and rigor and to
the extent current understanding allows questions such as: how and
to what extent do we understand the state of this system?; is it in
equilibrium?; is it hydrodynamized and locally isotropic?; what do
we know about its transport properties? 

As explained earlier, we know that at its peak the energy density
of the system is far in excess of that of hadrons, let alone nuclei.
There is no way that the system could be a tightly packed collection
of hadrons. Instead, it has to be described in terms of the quarks
and gluons themselves. The interplay between two crucial features
of QCD determines the nature of this state of matter. First, because
of asymptotic freedom and the high energies probed at RHIC and the
LHC it could be that the interactions between the quarks and gluons
are so weak that an equilibrium thermal state of matter would never
be reached. Second, at energy scales within an order of magnitude
of the confinement/deconfinement energy scale, QCD is strongly coupled.
The implication of this was not fully realized before experiments
at RHIC began \cite{Rischke:2005ne,QuarkGluonPlasmaRHIC}, as the
most common expectation was the formation of an equilibrated gas of
quarks and gluons with a temperature somewhat above the confinement/deconfinement
scale. We now realize that in this temperature range QCD describes
a relativistic fluid consisting of quarks and gluons that are so strongly
coupled to their neighbors that the resulting liquid cannot even be
described in terms of quasi-particles. The weak coupling picture must
be correct at early times in collisions with exceedingly high energy;
even in these collisions, the strong coupling picture would become
applicable later after a hydrodynamic fluid has formed. The question
of for how long during the initial moments of a RHIC or LHC collision
a weakly coupled picture can be applied remains open.

The crucial distinction between both scenarios can be found by measuring
the anisotropy of particles produced in heavy ion collisions. Qualitatively
this is easy to understand, as we saw in Section \ref{sec:Phenomenology-of-heavy}:
in the case of weakly interacting gas of particles, scatterings are
rare, the directions of the momenta of the gas particles are random,
the initial spatial anisotropy in the collision zone is washed out
by random motion, and the azimuthal distribution of particles in the
final state ends up isotropic. In this case, the measured two-particle
correlations are trivial, coming only from effects like momentum conservation
in late-time decays of hadrons. Alternatively, if the quarks and gluons
form a strongly coupled liquid soon enough, while the distribution
of energy density produced in the collision remains anisotropic, this
non-circular and lumpy drop of fluid will expand in a hydrodynamic
fashion, yielding faster expansion in the direction of larger gradients:
\textbf{hydrodynamics converts spatial anisotropies into momentum
anisotropy}. For perfectly circular collisions this would not lead
to an interesting distinction, but in the hydrodynamic picture we
would expect anisotropy arising because the incident nuclei are made
of nucleons and hence lumpy as well as an increasing anisotropy in
the particle spectrum as we probe less central, less circular, collisions.\footnote{It is also worth noting that when people have modeled the bulk dynamics
of the matter produced in heavy ion collisions via a system of colliding
particles, fitting such models to empirical observations inevitably
requires unphysically large scattering cross-sections (for example
parton-parton inelastic scattering cross-sections 15 times larger
than in perturbative QCD \cite{Molnar:2001ux} or values of $\alpha_{s}$
as large as 0.6 \cite{Xu:2007ns} or unphysically short mean free
paths). For example, in both the BAMPS \cite{Xu:2007ns} and AMPT
\cite{Lin:2004en} approaches, the particles in the model have mean
free paths that are much shorter than their de Broglie wavelengths.
Although these approaches differ from hydrodynamics in detail, at
a qualitative level what is happening in these models is that interactions
in a particulate model are being dialed up to a sufficient degree
that the model describes a fluid with low specific viscosity. (This
has been shown explicitly for BAMPS \cite{Xu:2007ns}.)}

To quantify the measurement of the azimuthal momentum anisotropy,
we perform a Fourier transformation on the angular distribution of
(charged) hadrons in the final state of the collision \cite{Ackermann:2000tr},
which results in the anisotropic flow coefficients $\bar{v}_{n}$,
defined from
\begin{equation}
\frac{d\bar{N}}{d\varphi}=\frac{\bar{N}}{2\pi}\left(1+2\sum_{n=1}^{\infty}\bar{v}_{n}\cos(n(\varphi-\bar{\Psi}_{n}))\right),\label{eq:anisotropic-flow}
\end{equation}
where $\varphi$ is the angle in the transverse plane, $\bar{\Psi}_{n}$
are the event plane angles (the first angle where the $n$th harmonic
component has its maximum multiplicity), and $\bar{N}$ is the average
number of particles of interest per event. All these observables can
in principle be measured as a function of rapidity, centrality, transverse
momentum and, around mid-rapidity (in collider experiments), also
differentially for different particle species. The second to fourth
harmonics are shown as a function of centrality in Fig. \ref{fig:elliptic-flow}
(middle), as extracted from the 2-particle correlator with particles
separated by a large gap in rapidity\footnote{There are several ways to measure the $v_{n}$ found in Eq. (\ref{eq:anisotropic-flow}),
most notably via measuring correlations among 4, 6, 8 or more particles,
or via analyzing particles separated in rapidity. Both techniques
are designed to exclude `jet-like' correlations between nearby particles
that come from the same jet shower or nearly back-to-back correlations
from pairs of jets. We shall not review the by now quite sophisticated
methods for extracting the $v_{n}$ \cite{Ollitrault:2009ie}. We
shall also not review the dependence of the $v_{n}$ on transverse
momentum or on hadron species \cite{Adams:2005dq}, even though their
dependence on particle momentum and mass provide important evidence
in support of their origin from a single hydrodynamic fluid with a
common flow velocity, or their distribution around their average value
in each centrality class, which also support a consistent picture.
(See e.g. \cite{Adams:2003am,Chatrchyan:2013kba,Foka:2016vta})}. (We shall come back to the hydrodynamic curves shortly.)

As anticipated, the system before hadronization indeed requires a
full hydrodynamic simulation in order to generate the sizable anisotropies
found. Hydrodynamics is a gradient expansion, assuming that a fluid
is everywhere close to thermal equilibrium, but allowing for small
gradients in both temperature and velocity field. In ideal (0th order)
hydrodynamics these gradients are ignored, which by assumption gives
an isotropic plasma in the plasma's local restframe. For viscous (first
order) hydrodynamics the gradients lead to an anisotropic stress tensor
$T_{\mu\nu}$ according to
\begin{eqnarray}
T_{\mu\nu} & = & \varepsilon\,u_{\mu}u_{\nu}+p[\varepsilon]\Delta_{\mu\nu}-\eta[\varepsilon]\,\sigma_{\mu\nu}-\zeta[\varepsilon]\Delta_{\mu\nu}\nabla_{\mu}u^{\mu}+\mathcal{O}(\partial^{2}),\text{ where}\label{eq:hydro-constituive}\\
\sigma_{\mu\nu} & = & \Delta_{\mu\alpha}\Delta_{\nu\beta}(\nabla^{\alpha}u^{\beta}+\nabla^{\beta}u^{\alpha})-\frac{2}{3}\Delta_{\mu\nu}\Delta_{\alpha\beta}\nabla^{\alpha}u^{\beta},\label{eq:sigma}\\
\Delta_{\mu\nu} & = & g_{\mu\nu}+u_{\mu}u_{\nu},
\end{eqnarray}
where $\varepsilon$ is the energy density and $u_{\mu}$ the velocity
field, both depending on the full space-time coordinates. In the local
fluid rest frame where $u_{{\rm LRF}}^{\mu}=(1,0,0,0)$ the projector
is given by ${\rm \Delta_{{\rm LRF}}^{\mu\nu}=diag}(0,1,1,1)$, and
in any frame $\Delta_{\mu\nu}u^{\mu}=\Delta_{\mu\nu}u^{\nu}=0.$ The
first two terms in (\ref{eq:hydro-constituive}) are just ideal hydrodynamics,
whereby the stress-energy tensor is given by an isotropic fluid with
energy density $\varepsilon$ that is boosted with a velocity $u_{\mu}$.
This fluid has a pressure that is given by the equation of state $p[\varepsilon]$,
which is an input into hydrodynamics that depends on the microscopic
properties of the theory under consideration. For heavy ion collisions
this is the QCD equation of state, which is usually obtained from
lattice calculations like those of Fig. \ref{fig:lattice} \cite{Huovinen:2009yb}
(see however \cite{Pratt:2015zsa}). Lattice calculations are also
used to relate the energy density to the temperature.

Beyond ideal hydrodynamics one needs to include corrections proportional
to gradients and consistent with the symmetries present. For scale
invariant viscous relativistic hydrodynamics it turns out that the
only transport coefficient possible at first order in gradients is
the shear viscosity $\eta[\varepsilon]$, which accompanies the $\sigma_{\mu\nu}$
of (\ref{eq:sigma}), containing first derivatives of the fluid velocity.
Close to the deconfinement/confinement transition, QCD is definitely
not scale invariant, and there it is also necessary to include the
term proportional to the bulk viscosity $\zeta[\varepsilon]$. Just
like $p[\varepsilon]$, the viscosities depend on the microscopic
properties of the theory, but these transport properties are notoriously
difficult to determine from a lattice calculation because they describe
the (time-dependent) process by which small deviations from equilibrium
relax whereas what is calculated directly on the lattice is (time-independent)
derivatives of the equilibrium partition function. We will return
to the determination of transport properties shortly.

Hydrodynamic evolution follows from the conservation of the stress-energy
tensor after specifying the equation of state, the transport coefficients
and the energy and velocity profiles at an initial time\footnote{In practice, solving the equations of viscous hydrodynamics is a bit
more involved since when they are discretized they contain modes with
wavelengths of order the discretization scale that propagate faster
than light. These modes are unphysical and are outside the regime
of the hydrodynamic gradient expansion, but because they are acausal
they make the numerical scheme unstable. This makes it necessary in
practice to solve a version of 2nd order hydrodynamics and verify
that the choice of 2nd order terms does not much affect the final
results, as must be the case if the gradient expansion is under control.
(See for example \cite{Baier:2006gy,Heinz:2013th,Romatschke:2017ejr})
We also note that we shall only review the application of hydrodynamics
to collisions at LHC and top RHIC energies and, at these energies,
for production of QGP far from the fragmentation regions. Extending
such calculations outside these regions, as relevant for the exploration
of the QCD phase diagram via the RHIC Beam Energy Scan mentioned in
Section 2, requires extending (\ref{eq:hydro-constituive}) to incorporate
the time evolution of the conserved baryon number current. It is well-known
how to do this \cite{Shen:2017ruz}, but less is known about the QCD
equation of state and transport coefficients at nonzero baryon chemical
potential. There is also an additional complication in that when the
Lorentz contraction of the incident nuclei is only moderate the dynamics
is intrinsically 3-dimensional.}. In the hydrodynamic evolution equations, $\nabla_{\mu}T^{\mu\nu}=0$,
the shear viscosity arises in the combination $\eta/(\varepsilon+p)=\eta/(Ts)$,
which is proportional to the length scale over which momentum can
be transported in the fluid \cite{Romatschke:2017ejr}. At weak coupling,
when the hydrodynamic fluid is made up of quasiparticles with a well-defined
mean free path $\lambda_{{\rm mfp}}$, it can be shown that $\eta/(\varepsilon+p)\propto\lambda_{{\rm mfp}}$
meaning that $\eta/s\propto T\lambda_{{\rm mfp}}$ \cite{York:2008rr,Casalderrey:2014,Romatschke:2017ejr}.
In a strongly coupled fluid, $\eta/s$ is well-defined and small,
but quasiparticles with mean free paths cannot be defined since attempting
to do so would result in a $\lambda_{{\rm mfp}}$ comparable to or
smaller than the de Broglie wavelength $1/T$. Whether the fluid is
weakly or strongly coupled, $\eta$ arises in the hydrodynamic equations
in this combination and it is the specific viscosity $\eta/s$ that
controls how rapidly sound waves, shear stress, or gradients of any
sort introduced in the initial conditions are dissipated into heat,
meaning that it is this quantity that is ultimately constrained by
comparing hydrodynamic calculations to data. To proceed further, it
is necessary to model the initial energy and velocity profile at some
proper time $\tau_{0}$. Fortunately we find that the insights we
present do not depend strongly on simplifying assumptions that we
have to make to solve our equations. A simple model used is to take
two discs of heavily Lorentz contracted nuclei to collide at some
impact parameter $b$ and from this construct an initial energy profile
that follows the overlap of the two discs (the Glauber model) with
an overall amplitude as a free parameter. The velocity profile is
often taken to be zero in the transverse plane and, in the longitudinal
direction, the evolution is assumed to be boost invariant around the
collision point at $t=z=0$. Since this assumption implies that the
longitudinal velocity is given by $v_{z}=z/t$, this gives a simple
and convenient model for an expanding plasma where all physics just
depends on proper time $\tau=\sqrt{t^{2}-z^{2}}$ and the transverse
coordinates. In state of the art hydrodynamic calculations which do
not assume boost invariance, $v_{z}=z/t$ remains a good approximation
but the initial distribution of energy density does depend on rapidity.

Having specified the initial conditions and the hydrodynamic equations,
the latter via choosing $\eta/s$ and taking $p[\varepsilon]$ and
$T[\varepsilon]$ from lattice calculations, it is possible to start
a simulation of the hydrodynamic evolution of this putative hydrodynamic
quark-gluon plasma. This simulation evolves the hydrodynamic variables
describing an expanding and cooling droplet of matter forward in time
up to a `freeze-out' hypersurface in space-time where the fluid temperature
has dropped to a specified value of order the temperature $T_{c}$
where the crossover from QGP to hadrons occurs. At the freeze-out
hypersurface, the fluid is converted into a thermal distribution of
hadrons, conserving energy and momentum \cite{Cooper:1974mv}. Subsequent
evolution is described via a gas of hadrons, which interact with each
other as further expansion and cooling occurs until all scattering
ceases at a lower `kinetic freeze-out' temperature. The resultant
ratios between the numbers of different hadron species, single particle
spectra for various hadron species, and anisotropy coefficients $v_{n}$
can all be directly compared with experimental data.

In this model, in order to generate as much transverse flow (both
isotropic, or radial, flow and anisotropic flow as described by the
$v_{n}$) as seen in data it is necessary to take $\tau_{0}$ smaller
than $1.0$ fm/c, in some calculations as small as 0.2 fm/c. (In more
advanced models that include the growth of the transverse velocity
before $\tau_{0}$, this constraint can be somewhat weaker \cite{vanderSchee:2013pia}.)
The amplitude of the initial energy density profile for central collisions
($b=0$) is fitted to give the observed total particle multiplicity
per unit rapidity. The multiplicity as a function of impact parameter
$b$ is then a prediction of the model, which can be compared with
the experimental results and used to obtain a constraint on $\eta/s$
.

The precise magnitude of the anisotropies $v_{n}$ then depends quite
sensitively on the viscosity of the plasma. Already from the relatively
straightforward simulation with smooth initial conditions described
above, it can be estimated that $\eta/s\sim0.08-0.20$ during the
hydrodynamic phase in heavy ion collisions at RHIC energies \cite{Romatschke:2007mq}
(see \cite{Schenke:2010rr} for a full 3D hydrodynamic simulation).
This is one of the greatest discoveries of the heavy ion programs
at RHIC and LHC: the experimental data is well described by the hydrodynamic
evolution of a droplet of quark gluon plasma with a specific viscosity
smaller than that of any other fluid known in nature. Quark-gluon
plasma is hence sometimes referred to as the most perfect liquid. 

As an example, and to give a sense of how well the system is understood
currently, in Figure \ref{fig:elliptic-flow} (middle) a more precise
comparison between experimental data and one particular hydrodynamic
calculation is made. A crucial ingredient in this computation is the
initial condition for the (lumpy, fluctuating) transverse profile,
which is taken from a Monte Carlo Glauber model, with fluctuating
positions of individual protons and neutrons, convolved with fluctuations
of the energy density within a single nucleon that is based upon a
saturation model in which color fields are large in magnitude but
weakly coupled. Fluctuations in the initial state are necessary to
obtain agreement with current, precise, data including in particular
the $v_{n}$ anisotropies in head-on collisions and the odd harmonics
such as $v_{3}$ in Figure \ref{fig:elliptic-flow} (middle). Without
such fluctuations, the collisions would be perfectly symmetric under
parity in the $\vec{b}$ direction, which would imply that all harmonics
$v_{n}$ with $n$ odd would vanish. In actual collisions $v_{3}$
is in fact larger than the higher harmonics, showing that fluctuations
break this parity symmetry \cite{Alver:2010gr}. The authors of Ref.
\cite{Niemi:2015qia} evolved hydrodynamics with five different assumptions
for $\eta/s$ as a function of temperature, see Figure \ref{fig:elliptic-flow}
(middle). Current analyses of data are beginning to yield some constraints
on this temperature-dependence. It is also possible to introduce a
bulk viscosity at temperatures near the QCD phase transition \cite{Ryu:2015vwa},
where it is expected to be important. The bulk viscosity in \cite{Ryu:2015vwa}
is needed to get an accurate fit of the transverse momentum spectrum.
The authors of \cite{Ryu:2015vwa} find that introducing bulk viscosity
improves the fit to single particle transverse momentum spectra without
spoiling the quality of the fit for the elliptic flow $v_{2}$, but
the optimal value of $\eta/s$ for the matter produced in LHC heavy
ion collisions changes from $0.16$ to $0.095$. 

In addition to providing a more accurate description of the systematic
dependences of the measured flow coefficients upon averaging, an event-by-event
analysis of a large ensemble of events with fluctuating initial conditions
also makes it possible to compare hydrodynamic calculations of the
distributions of the $v_{n}$ coefficients to experimental distributions.
It then turns out that the distribution $\delta v_{n}\equiv(v_{n}-\langle v_{n}\rangle)/\langle v_{n}\rangle$
is largely independent of the hydrodynamic transport coefficients
but is instead sensitive to the initial shape of the energy density
\cite{Niemi:2012aj} (Fig. \ref{fig:elliptic-flow}, right), including
its lumpiness. Hence, these distributions are an excellent way to
constrain the hydrodynamic initial conditions, after which other observables
can then be used with greater confidence to constrain transport coefficients
such as $\eta/s$. The correlations between different event plane
angles $\Psi_{n}$ also turn out to be a useful event-by-event observable.
These correlations are not only sensitive to the average $\eta/s$
during the hydrodynamic evolution, but can also begin to constrain
different hypotheses for the temperature dependence of $\eta/s$ \cite{Niemi:2015qia}.

It is an essential question how the conclusions about the shear viscosity
depend on the model used, especially considering the uncertainty in
the initial profiles as well as in the bulk viscosity and the temperature
dependence of the shear viscosity. It has recently become possible
to study the model dependence more systematically by doing a Bayesian
analysis over a space of model parameters that include most models
available, with recent estimates obtained via fitting to many different
kinds of data from both RHIC and the LHC giving $\eta/s\approx(0.07_{-0.04}^{+0.05})+c(T-T_{c})$
\cite{Bernhard:2016tnd} for temperatures $T>T_{c}=154$ MeV with
$T_{c}$ corresponding to the crossover between QGP and hadrons, but
where the constant $c$ is at present only constrained to be between
0 and 1.58/GeV. 

The small value for the dimensionless shear viscosity ratio $\eta/s$
is especially interesting. At weak coupling, this ratio is proportional
to the ratio of the quasiparticle mean free path to the mean spacing
between quasiparticles. A larger value of the ratio means that momentum
can more easily be transported over significant distances, which is
what is required in order to dissipate shear stress into heat. And,
weaker coupling means larger values of this ratio. (For a weakly coupled
gas of gluons, in fact $\eta/s\sim1/g^{4}\log(1/g)$\cite{Arnold:2000dr},
where $g$ is the QCD coupling, namely the QCD analogue of $e$ in
electromagnetism.) At strong coupling, on the other hand, each volume
element of the QGP fluid is so strongly coupled to its neighbors that
very little (net) momentum can be transferred to nearby fluid elements,
meaning that velocity gradients remain, shear stress does not dissipate,
and the specific viscosity is small. The measured value of $\eta/s$
for QGP turns out to be so small, however, that the fluid cannot be
described in terms of quasiparticles with mean free paths since to
do so would require mean free paths that are smaller than $1/T$.
Strikingly, for certain infinitely strongly coupled quantum theories
with a large number of degrees of freedom that are described by a
holographic dual gravitation theory, it can be computed that $\eta/s=1/4\pi\approx0.08$
\cite{Policastro:2001yc}, which is conspicuously close to the (average)
viscosity found in hydrodynamic calculations used to model the dynamics
of droplets of QGP produced in heavy ion collisions. Although QCD
itself is not known to have a holographic dual, this motivates using
gauge theories which do have dual gravitational descriptions to model
dynamics in heavy ion collisions, as we shall elaborate later. 

\section{Thermalization, Hydrodynamization and Isotropization}

\begin{figure}
\begin{centering}
\begin{minipage}[t]{0.56\columnwidth}%
\begin{center}
\includegraphics[width=1\textwidth]{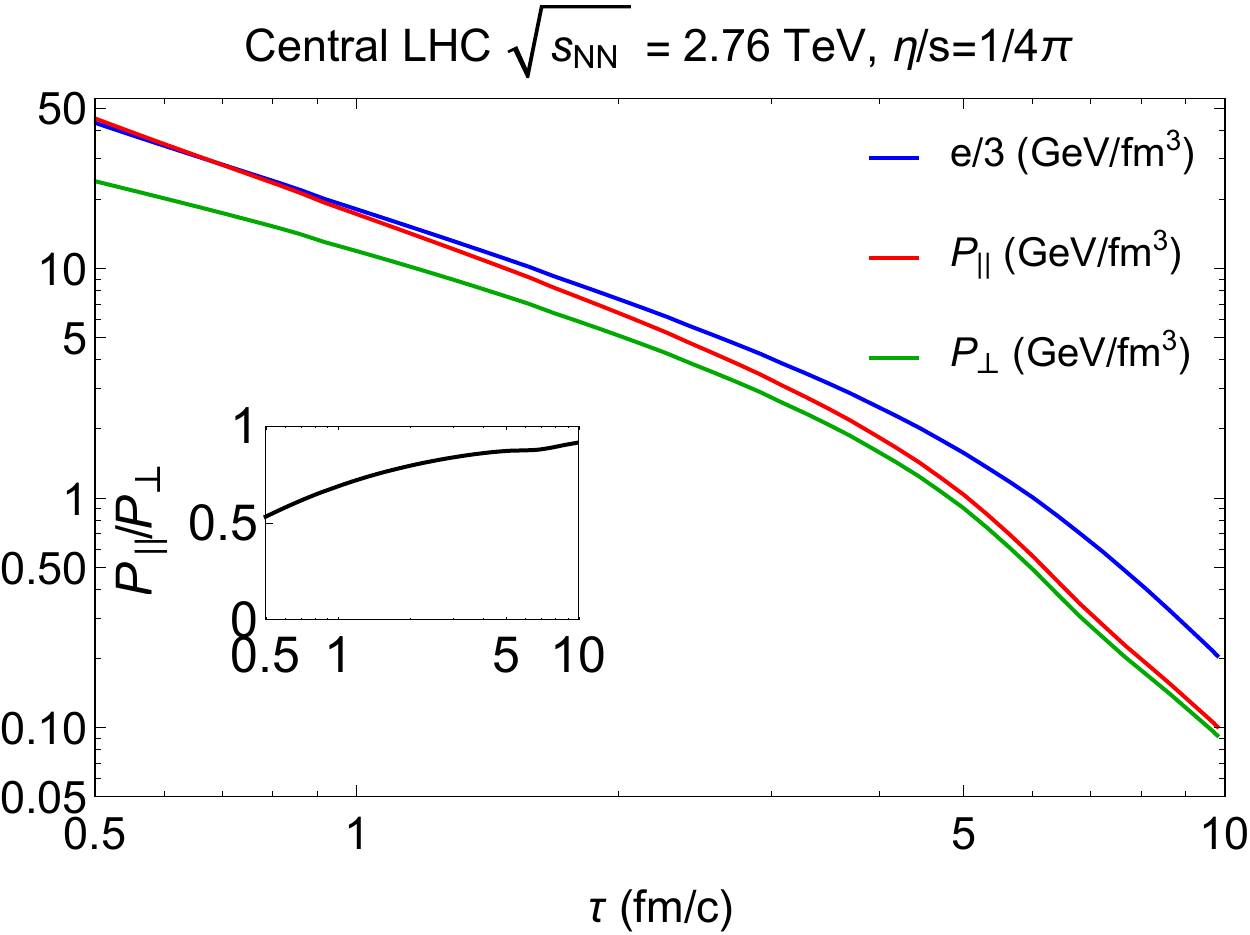}
\par\end{center}%
\end{minipage}
\par\end{centering}
\caption{Typical energy density $e$, longitudinal pressure $p_{||}$, and
transverse pressure $p_{\perp}$ as a function of proper time $\tau$
for a central collision at LHC, at the center of the transverse plane,
with the pressure anisotropy as an inset. (Figure adapted from \cite{vanderSchee:2013pia})
\label{fig:energy-and-pressure}}
\end{figure}

The success of the hydrodynamic paradigm begs the question: when,
why and how does the colliding debris begin to be accurately described
by hydrodynamics, which is to say \textbf{hydrodynamize}? We would
like to understand the underlying physics behind these questions,
preferably from a QCD point of view. A related question is whether,
if hydrodynamics is applicable for most of the evolution, this implies
that the QGP formed in a heavy ion collision also thermalizes fast.
The answer need not be yes, since a thermal state, by definition,
is locally isotropic and free of gradients whereas hydrodynamics can
work well when contributions that are first order in gradients are
significant while those coming from higher order gradients are small.
(Also note recent work \cite{Heller:2015dha,Romatschke:2017ejr,Denicol:2017lxn}
that succeeded in resumming all orders in gradients in a simplified
geometry.) We hence have to ask how the pressure anisotropy evolves
during the hydrodynamic evolution.

The droplet of QGP formed in a heavy ion collision is expanding rapidly,
and even though it has the smallest dimensionless specific viscosity
$\eta/s$ found in nature, the gradients present due to the fast expansion
(initially mostly in the longitudinal direction) imply that the viscous
corrections, which are first order in gradients, are in fact sizable.
Indeed, when $v_{z}=z/t$, as for a boost-invariant velocity profile,
it is clear that at early times the gradients of the velocity field
are large. When extracting the pressure anisotropy as in Fig. \ref{fig:energy-and-pressure}
it can be seen that the gradient corrections are important for proper
times between $0.3$ and $6$ fm/c and the plasma only becomes approximately
isotropic after a proper time of roughly $\tau=6$ fm/c \cite{vanderSchee:2013pia}. 

So even if hydrodynamics is indeed a good description around times
as early as $0.5$ fm/c, because of the significant initial gradients
and the smallness of the specific viscosity, the fluid does not fully
isotropize and hence thermalize before a much later time of around
$6$ fm. We say that the fluid hydrodynamizes rapidly, within a proper
time of around $0.2-0.6$ fm/c or $0.4-1.0$ fm/c at LHC or RHIC energies
respectively (appropriately, these are typical starting times used
in the hydrodynamic simulations described in Section \ref{sec:A-hydrodynamic-fluid}),
with this hydrodynamization followed subsequently by an extended period
of hydrodynamic evolution with significant gradients in the fluid,
before isotropization and complete thermalization at a substantially
later time. 

The estimates of the hydrodynamization time that we have quoted are
often based upon assuming that when QGP hydrodynamizes in a heavy
ion collision it does so without any initial transverse fluid velocity.
This extra assumption is, however, unnecessary and in fact we will
see shortly that all theoretical frameworks would predict the generation
of some transverse flow already during the far-from-equilibrium, pre-hydrodynamization,
stage of the collision, which can hence resemble hydrodynamic evolution.
The question of when the QGP formed in a heavy ion collision hydrodynamizes
is hence intricately linked to how hydrodynamics becomes applicable,
and in particular how much the far-from-equilibrium pre-hydrodynamic
dynamics resembles hydrodynamics itself. We will return to this in
the next section.

\section{Initial stage\label{sec:Initial-stage}}

\begin{figure}
\begin{centering}
\includegraphics[width=1.2\textwidth]{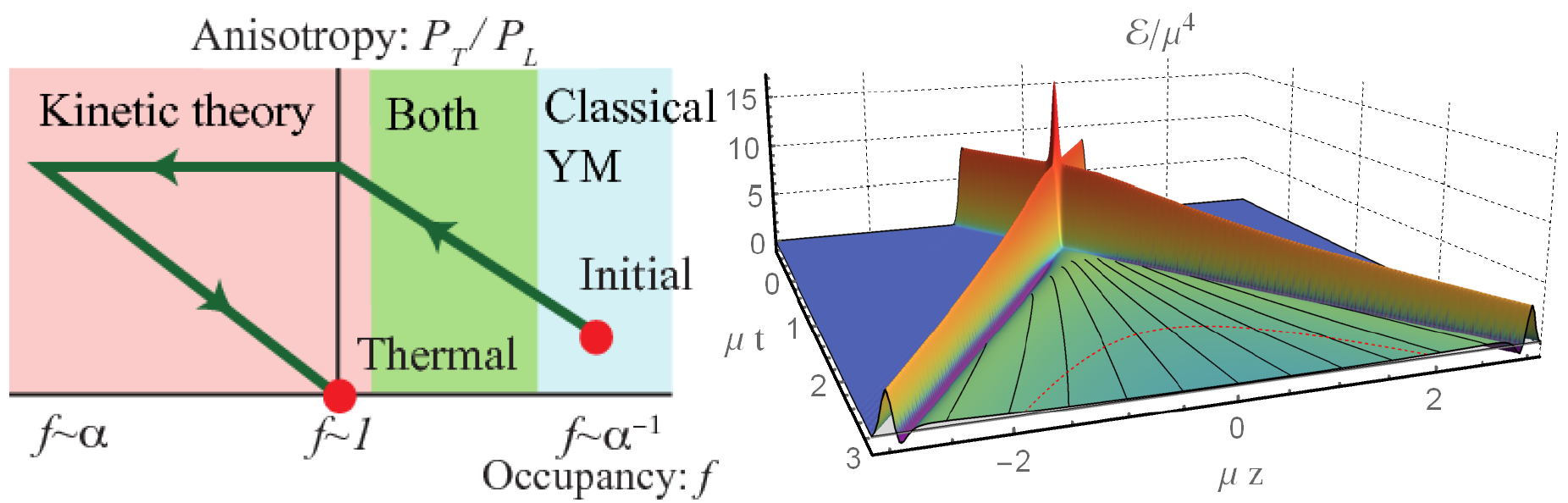}
\par\end{centering}
\caption{(left) We provide a cartoon of how the pressure anisotropy (vertical
axis) and $f$, the typical occupation number of modes with momentum
$\sim Q_{s}$ (see text) in the gluon wave function, evolve during
the initial stages of a heavy ion collision if one assumes that this
can be described entirely at weak coupling. The weakly coupled dynamics
can be described in terms of classical Yang-Mills fields if $f\gg1$
and in terms of kinetic theory if $f\ll1/\alpha_{s}$, meaning that
as long as $\alpha_{s}$ is small enough there is a regime in which
both descriptions are valid. The evolution begins with classical gluon
fields at high occupancy $f\sim1/\alpha_{s}$ and moderate pressure
anisotropy, proceeds to kinetic theory at low occupancy and large
pressure anisotropy, after which the matter thermalizes, meaning that
the pressures become isotropic and the occupancy of modes in the kinetic
theory reaches $f\sim1$. (right) The plot shows the energy density
as a function of time and longitudinal coordinate $z$ in units of
$\mu$ for a collision of two highly contracted parallel sheets of
energy in strongly coupled SYM theory colliding head-on along the
$z$-direction at $t=z=0$. Here, $N_{c}^{2}\mu^{3}/2\pi^{2}$ is
the energy per transverse area of the incident sheets of energy, with
$N_{c}$ the number of colors in the theory. The energy density contains
far-from-equilibrium regions, even including regions where energy
density is negative and a restframe cannot even be defined \cite{Arnold:2014jva}.
After the red dashed line the evolution of the plasma (in green) is
hydrodynamic within 5\% accuracy \cite{Chesler:2015fpa} (Figs. from
\cite{Kurkela:2016vts,Casalderrey-Solana:2013aba}). \label{fig:initial-stage}}
\end{figure}

The hydrodynamic model described above works well, perhaps surprisingly
well, explaining many features of the particle spectra and the anisotropic
flow coefficients. This poses three urgent questions: How does the
debris left after a heavy ion collision evolve into an almost perfect
hydrodynamic fluid \textbf{so fast}? And, how should this initial
non-hydrodynamic stage in the dynamical evolution be described and
in what initial conditions for the hydrodynamic stage does this result?
We shall sketch the present understanding of both questions. The third
question is how is entropy produced? This question provides a further
reason for interest in the initial stage because almost all of the
entropy produced in a heavy ion collision is produced before hydrodynamization:
because the specific viscosity of the hydrodynamic liquid is so small,
very little additional entropy is produced during the later, longer,
hydrodynamic expansion. This means that the multiplicity of particles
produced in the final state of a heavy ion collision is controlled
by the dynamics occurring during its initial stage.

From a purely QCD point of view these questions are unfortunately
hard to answer because non-perturbative real time dynamics cannot
be studied on the lattice. During the initial stage, and in particular
during its earliest moments, many of the important scattering processes
involve high transverse momentum transfer and hence can be described
using pQCD. Soft, strongly coupled interactions are also important,
in particular later during the initial stage as the matter hydrodynamizes,
namely as it is becoming a strongly coupled fluid. It is therefore
reasonable that various authors have developed entirely weak coupling
descriptions of the initial stage while at the same time various authors
have modeled far-from-equilibrium dynamics non-perturbatively using
holography. We consider the two approaches in turn.

In the context of pQCD, the starting point for the description of
the initial stages involves the phenomenon of ``saturation'' in
the gluon wave function of the incident nuclei \cite{McLerran:1993ni}.
When colliding ions at higher and higher energies, the gluons that
collide and end up near mid-rapidity after the collision are gluons
from the parton distribution function (PDF) of the incident nuclei
with smaller and smaller momentum fraction $x=p_{z}/P$, defined with
respect to the momentum of the nucleon $P$. In a perturbative analysis,
at large $Y=\log(1/x),$ the gluon PDF increases rapidly with increasing
$Y$. At mid-rapidity in collisions at higher and higher energy, meaning
smaller and smaller $x$, there will be more and more gluons. Until,
that is, above some gluon density, gluon merging becomes as important
as gluon splitting as $x$ is decreased further. The occupation number
of gluon modes in momentum space with this value of $x$ and below
is of order $1/\alpha_{s}$ and this component of the wave function
of the nucleus is referred to as saturated. The typical transverse
momentum of these saturated gluons is referred to as the saturation
scale $Q_{s}$ and the number density of these gluons per unit area
in the transverse plane is then given by $Q_{s}^{2}/\alpha_{s}$.
The premise of the quantitative version of this analysis is that $\sqrt{\alpha_{s}(Q_{s})}$
(where $\alpha_{s}$ is the running QCD coupling constant which becomes
small at high momentum transfer) is small when evaluated at the scale
$Q_{s}$. It is at the saturation scale where we find the low-$x$
gluons which dominate the interaction in an ultrarelativistic heavy
ion collision. (See \cite{Iancu:2012xa} for an accessible introduction.)

The above perturbative analysis implies that just after a heavy ion
collision one ends up at mid-rapidity with gluon modes with transverse
momenta up to $\sim Q_{s}$ that are over-occupied. Making this analysis
more quantitative leads to the conclusion that $Q_{s}$ is of order
$1$ or $3$ GeV for collisions at RHIC or the LHC, not so high as
to make the assumptions of the perturbative treatment incontrovertible.
Next, these gluons with transverse momenta of order $Q_{s}$ radiate
softer gluons and scatter with the growing bath of softer gluons until
hydrodynamization is achieved \cite{Baier:2000sb,Baier:2002bt}. These
processes are somewhat involved, and can be described via weakly coupled
classical field theory or an effective kinetic theory of weakly coupled
partons in overlapping regions of parameter space. (The first can
be used when there are modes with occupation numbers that are $\gg1$;
the second works for occupation numbers that are smaller than $1/\alpha_{s}$.)
Plasma instabilities can play a role in the classical approach, although
to leading order the classical evolution is self-similar due to the
rapid longitudinal expansion \cite{Berges:2013eia}. This expansion
also drives the occupation numbers down, though, and at later times
during the pre-hydrodynamic stage the effective kinetic theory must
be used. (See Fig. \ref{fig:initial-stage}.) The earliest analyses
of these processes yielded the conclusion that in the limit of very
weak coupling the parametric dependence of the hydrodynamization time
is $\tau_{{\rm hydro}}\gtrsim\alpha_{s}^{-13/5}Q_{s}^{-1}$ \cite{Baier:2000sb,Baier:2002bt}.
As numerical analyses of both the classical and the kinetic evolution
have advanced, the currently most quantitative estimate is that, if
one assumes $\alpha_{s}=0.3$,\begin{marginnote}[] \entry{'t Hooft coupling}{The smallness of $\alpha_s$ controls perturbative corrections in QCD at weak coupling while the largeness of the 't Hooft coupling $\lambda \equiv 4\pi \alpha_s N_c$ controls finite coupling corrections in holography.  $\alpha_s = 0.3$ corresponds to  $\lambda \approx 11$, meaning that this coupling is neither close to nor far from both 0 (perturbative methods) and  $\infty$ (holography).} \end{marginnote}
the kinetic theory description of the energy density, transverse pressure,
and longitudinal pressure hydrodynamizes after a time that is about,
or even a little less than, $1$ fm/c \cite{Kurkela:2015qoa}.

We know the specific viscosity is small and the coupling strong in
the hydrodynamic liquid. This motivates exploring strongly coupled
analyses of hydrodynamization as an alternative path to insights.
The option that has been pursued most successfully is to analyze the
complete far-from-equilibrium initial stage assuming that the dynamics
is strongly coupled throughout using holography, which provides a
dual gravitational description for certain gauge theories around infinitely
strong coupling \cite{Casalderrey:2014}. This duality is truly remarkable,
as it maps intractable real-time far-from-equilibrium non-perturbative
QFT problems onto equivalent, but tractable, computations within classical
general relativity in Anti-de Sitter space, a (4+1)-dimensional space-time
with a negative cosmological constant. Due to the strong interactions,
the hydrodynamization time can be much shorter than at weak coupling.
An early hint of this was the discovery that small perturbations around
an equilibrium thermal state (equivalent to exciting quasi-normal
modes of the dual black hole horizon) relax exponentially with characteristic
time $\tau\sim1/\pi T$ \cite{Horowitz:1999jd}. Computations of the
relaxation of many far-from-equilibrium disturbances to boost-invariant
expanding flows \cite{Chesler:2009cy,Heller:2011ju} have shown that
hydrodynamization occurs within a time $\tau_{{\rm hydro}}\sim0.7/T_{{\rm hydro}}$,
where $T_{{\rm hydro}}$ is the temperature at which hydrodynamization
occurs, and furthermore show a remarkably wide applicability of the
quasi-normal mode analysis \cite{Heller:2012km}.

\begin{textbox}[t]\section{Holography} started with a seminal paper by Maldacena \cite{Maldacena:1997zz}, which provides an exact equivalence between certain string theories and certain (supersymmetric) gauge theories. In one direction, this exact equivalence has led to a much better understanding of quantum gravity by using gauge theory dynamics. To use the equivalence in the other direction, it is also possible to take the limit where string theory becomes a theory of ordinary classical gravity in a curved space-time with a negative cosmological constant and one extra dimension. In that case, the gauge theory has many colors and is infinitely strongly coupled.  The equivalence can then provide reliable insights into complex dynamical questions in a  strongly coupled gauge theory. Position in the extra dimension encodes the length scale of excitations in the gauge theory. For example, the position of a horizon in the gravitational spacetime  corresponds to $1/T$ in the gauge theory, with $T$ the temperature of the strongly coupled plasma with $\eta/s=1/4\pi$ \cite{Policastro:2001yc,Casalderrey:2014}. Because all aspects of a one-higher-dimensional gravitational theory are encoded in features of the gauge theory, the mapping is referred to as a holographic duality. \end{textbox} 

More advanced calculations permit the complete and rigorous simulation
of the collision of sheets or discs of energy density in the infinitely
strongly coupled super-Yang-Mills theory that is a cousin of QCD with
a dual holographic description \cite{Chesler:2010bi,Casalderrey-Solana:2013aba}
from the moment of collision through hydrodynamization and subsequent
hydrodynamic expansion and cooling, including the development of radial
and elliptic flow \cite{Chesler:2015wra,vanderSchee:2012qj}. This
allows for direct and quantitative analyses of the hydrodynamization
process after a collision, analyses which yield an affirmation of
the hydrodynamic picture sketched above. In this context, the most
important conclusion is that a system that begins with an ultrarelativistic
collision can become hydrodynamic quickly, with a collisions starting
from a wide range of initial conditions yielding values for $\tau_{{\rm hydro}}T_{{\rm hydro}}$
between 1/4 and 1, as well as a hydrodynamic fluid that is initially
strongly anisotropic, with significant gradients. For the hydrodynamic
calculation of Fig. \ref{fig:energy-and-pressure}, solving the equation
$\tau\,T(\tau)=1$ leads to $\tau_{{\rm hydro}}\approx0.35$ fm/c,
whereby at that time $T_{{\rm hydro}}\approx560$ MeV. Hydrodynamization
may occur at an even earlier time and hotter temperature if $\tau_{{\rm hydro}}T_{{\rm hydro}}\lesssim1$.

These calculations yield other qualitative insights about the pre-hydrodynamic
stage in a collision. For example, they show that the far-from-equilibrium
dynamics of the collision yields a hydrodynamic fluid whose longitudinal
velocity profile is to a very good approximation boost invariant but
whose energy (and entropy) density profile is far from boost invariant,
taking on a shape that is approximately Gaussian in rapidity with
a width of 0.98 \cite{Chesler:2015fpa}. This is qualitatively in
line with what is seen empirically, but is too narrow. Calculations
have also been performed that follow the collision of strongly coupled
sheets of energy density that carry ``baryon number'' (a conserved
quantum number introduced by hand in the holographic gauge theory)
showing that after the collision the ``baryon number'' distribution
is also centered on mid-rapidity \cite{Casalderrey-Solana:2016xfq},
rather than losing only a few units of rapidity as in QCD. This, and
the narrowness of the energy/entropy distribution, are almost certainly
consequences of the fact that the gauge theory used in these calculations
is not asymptotically free. The fact that in QCD the coupling is weak
at the earliest moments of the collision is indeed important. This
provides strong motivation for recent developments in the holographic
framework, including collisions in theories that are not scale invariant
\cite{Attems:2016tby} and that feature weaker-than-infinitely-strong
coupling \cite{Grozdanov:2016zjj,Waeber:2015oka} which give a shear
viscosity that is larger than canonical, a nonzero bulk viscosity,
and somewhat larger hydrodynamization times. It will be quite interesting
to see how the distributions of energy, entropy and ``baryon number''
change in these collisions. As a final example, these calculations
permit the assessment of how much radial transverse flow develops
already before hydrodynamization. While hydrodynamic gradients of
course generate this flow later, early far-from-equilibrium evolution
can do so too, and in fact at strong coupling it is found that more
pre-hydrodynamic flow is generated than would arise if this earliest
epoch were instead hydrodynamic \cite{vanderSchee:2012qj,Habich:2014jna}.
Similar results have also been obtained in an equivalent study at
weak coupling \cite{Keegan:2016cpi}. 

A question that holographic calculations have not yet addressed (because
to date they have not included any representation of the fact that
in QCD the Lorentz contracted incident nuclei are made of nucleons)
is how the lumpiness of the energy density is distributed over the
transverse plane at the start of the hydrodynamic stage in a heavy
ion collision. In order to make comparisons to the increasingly precise
measurements of azimuthal anisotropies and correlations described
in Section \ref{sec:Phenomenology-of-heavy}, the fluctuations in
the energy density across the transverse plane must be included. All
phenomenological modeling includes the lumpiness coming from the initial
positions of the participating (or ``wounded'') nucleons inside
the colliding nuclei, via the Monte Carlo Glauber model described
in Section \ref{sec:Phenomenology-of-heavy}. The simplest models
just assume that each wounded nucleon contributes a Gaussian blob
of energy density, but the precision of present data is sufficient
that fluctuations that are somewhat smaller than a nucleon must be
included in order to optimize model predictions. Refined models translate
the density of wounded nucleons into a locally varying saturation
scale, and then use this as a guide to placing fluctuating color sources
in the transverse plane, sources which in turn drive the numerical
evolution of classical Yang-Mills fields whose stress-energy tensor
is then used to initialize hydrodynamics \cite{Schenke:2012wb,Gale:2012rq}.
Much remains to be done, including implementing an intermediate kinetic
theory description, introducing lumpiness into the holographic calculations
to provide a strong coupling benchmark, and in the long run testing
the predictions of saturation calculations for the gluon distribution
across the transverse plane in the incident nuclei against measurements
at a future electron-ion collider \cite{Accardi:2012qut}. However,
the best available calculations that begin with an initially lumpy
energy density and follow its hydrodynamic evolution give an excellent
simultaneous fit for RHIC and LHC to the probability distributions
of the $v_{n}$'s even for very off-central collisions \cite{Niemi:2015qia}.

\section{Jets in quark-gluon plasma\label{sec:Jets-in-quark-gluon}}

In occasional heavy ion collisions, partons from the incident nuclei
scatter off each other at very large momentum transfer, creating two
or more quarks or gauge bosons with very high transverse momentum
(many tens of GeV at RHIC; as high as 100 or even 1000 GeV at the
LHC). When such a hard scattering occurs in a proton-proton collision,
each hard parton that is produced showers into a spray of softer partons
within some irregular cone in momentum space, called a jet. Jet production
and showering in vacuum is well described by perturbative QCD \cite{Sjostrand:2007gs}.
When a jet is produced in a heavy ion collision, the partons in the
shower must plow through the droplet of QGP produced in the same collision.
As this happens, the jet partons: (i) lose energy and forward momentum,
(ii) pick up momentum transverse to their original direction, and
(iii) deposit energy and momentum into the droplet of QGP, creating
a wake. The first of these phenomena is well-established experimentally
\cite{Chatrchyan:2012nia,ATLAS:2017wvp} and there are strong indications
of the third \cite{Khachatryan:2015lha,CMSsoft}. The second, which
is referred to as momentum broadening since it can broaden the shape
of a jet in momentum space, is apparent in all theoretical approaches
but has not yet been seen experimentally \cite{Chatrchyan:2012nia,Casalderrey-Solana:2016jvj}. 

In the longer term, and in particular once we have high statistics
jet data at RHIC from the future sPHENIX detector \cite{Adare:2015kwa}
and from higher luminosity running at the LHC in the early 2020s,
the motivation for precision analyses of how jets are modified via
their passage through QGP is that this may teach us about the inner
workings of QGP. This is the closest we can ever come to probing QGP
by doing a scattering experiment and, as we discussed in Section \ref{sec:Why-do-we},
this is the best possible path toward addressing one of the big open
questions in the field: how does a strongly coupled liquid emerge
from an asymptotically free gauge theory? When the short-distance
structure of QGP is resolved, it must consist of weakly coupled quarks
and gluons. And yet, at length scales of order $1/T$ and longer they
become so strongly correlated as to form a liquid. Just as Rutherford
found nuclei within atoms and Friedman, Kendall and Taylor found quarks
within protons by doing scattering experiments, in the longer term
experimentalists hope to see the short-distance particulate structure
of QGP by seeing rare events in which a jet parton resolves, and scatters
off, a parton in a droplet of QGP.

\begin{figure}
\begin{centering}
\includegraphics[width=1.2\textwidth]{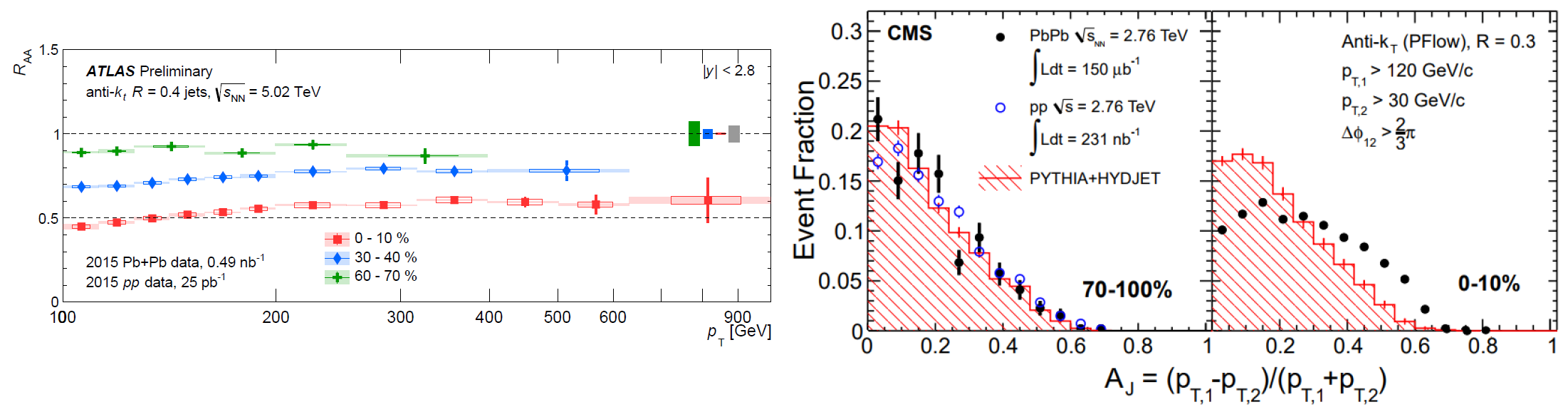}
\par\end{centering}
\caption{On the left we show the nuclear modification factor $R_{AA}$ for
jets for three different centralities as a function of jet transverse
momentum $p_{T}$ \cite{ATLAS:2017wvp}. On the right we show the
dijet asymmetry $A_{J}$ for $pp$ collisions and for peripheral (left)
and central (right) heavy ion collisions. The PYTHIA+HYDJET distribution
shows the expected asymmetry if no nuclear effects were present \cite{Chatrchyan:2012nia}.\label{fig:RAA}}
\end{figure}

There are many physics questions (involving larger or more common
effects) that are very interesting in their own right that must be
understood quantitatively before realizing the vision of using jets
as microscopes trained upon a droplet of QGP. This program is well
underway, and could easily be the subject of an entire review of its
own (see e.g. \cite{Mehtar-Tani:2013pia,Qin:2015srf,Connors:2017ptx}).
The most basic observation is that jets lose a substantial amount
of energy, often 10 GeV or more, as they traverse a droplet of QGP.
Noting that losing this amount of energy over only a few fm of distance
corresponds to an enormous $dE/dx$, this provides a direct, and completely
independent, confirmation that the matter produced in a heavy ion
collision is strongly coupled. This energy loss can be seen in many
ways including for example just by counting the number of jets with
a given (high) transverse momentum: it is suppressed in heavy ion
collisions relative to what would be seen in $N_{{\rm coll}}$ $pp$
collisions, which is to say relative to the expected number of jets
if there would be no interaction with the medium as explained in Section
\ref{sec:Phenomenology-of-heavy}. This is quantified by the nuclear
modification factor \cite{Adcox:2001jp}\begin{marginnote}[] \entry{Nuclear modification factor}{Ratio of the number of some countable objects  (e.g.  jets defined via a specified reconstruction procedure with a given $p_T$, hadrons of a specified type with a given $p_T$, etc.) found in nuclear collisions divided by the (theoretical) value that would be expected from an analogous number of proton-proton collisions without the presence of a medium.} \end{marginnote}
\begin{equation}
R_{AA}(p_{T})=\frac{dN^{AA}/dp_{T}}{\langle N_{{\rm coll}}\rangle dN^{pp}/dp_{T}},
\end{equation}
with $dN^{xx}/dp_{T}$ the number of jets (or in other contexts particles
of a specified type) produced in $AA$ or $pp$ collisions. Indeed,
Fig. \ref{fig:RAA} shows a large suppression of these jets, especially
for central collisions in which the droplet of QGP that the jets need
to traverse is the largest. A crucial check of this procedure is the
fact that high $p_{T}$ colorless probes, such as $\gamma$'s or Z-bosons
are indeed found to have $R_{AA}=1$, as expected since they do not
interact with QGP 

Throughout the study of jets it is important to realize that jets
with a high transverse momentum $p_{T}$ are produced with a probability
that drops very rapidly with increasing $p_{T}.$ The production probability
for jets produced at mid-rapidity with values of $p_{T}$ that are
not within an order of magnitude of the beam energy scales roughly
as $p_{T}^{-6}$ \cite{Spousta:2015fca}. The steepness of the energy
spectrum implies that a small fractional jet energy loss corresponds
to a large suppression in $R_{{\rm AA}}$ for jets. (As a contrafactual
example, if we imagine that all jets lose 10\% of their energy, i.e.
jets with $100$ GeV started as $110$ GeV jets, then since $110$
GeV jets are approximately $(100/110)^{6}\approx56\%$ rarer than
$100$ GeV jets it follows that we would observe a nuclear modification
factor of $R_{AA}\approx0.56$.) In reality, different jets with the
same initial energy lose very different amounts of energy as we shall
describe below, meaning that this argument must be made at the ensemble
level, but the conclusion is the same: because of the steepness of
the jet energy spectrum the suppression in $R_{{\rm AA}}$ for jets
is a sensitive measure of jet energy loss. Note that this argument
does not apply in the same way to $R_{{\rm AA}}$ for high-$p_{T}$
hadrons, as we shall see below.

Figure \ref{fig:RAA} (right) illustrates another way of seeing that
jets lose energy, and also provides direct evidence that in a given
event some jets lose more energy than others. This arises for two
reasons. First, the characteristics of jets with a given energy vary
quite considerably and there are now a variety of theoretical arguments
(at both weak and strong coupling) that indicate that a jet that fills
a cone with a wide opening angle (and at weak coupling contains many
partons) loses much more energy than a narrower jet with the same
energy carried by fewer harder partons \cite{Milhano:2015mng,Chesler:2015nqz,Rajagopal:2016uip,Casalderrey-Solana:2016jvj}.
Because of the steepness of the jet spectrum described above, the
ensemble of jets that comes out of the droplet of QGP will be dominated
by those jets that lost relatively little energy, meaning that the
jets that survive in a heavy ion collision with a given energy are
likely to be those that started out the narrowest and are on average
narrower than typical jets with the same energy in proton-proton collisions.
There is some evidence for this effect in measured jet shapes \cite{Khachatryan:2016tfj}.
Note that measuring $R_{{\rm AA}}$ for high-$p_{T}$ hadrons is quite
different: in both $pp$ and $AA$ collisions, a high-$p_{T}$ hadron
is statistically likely to come from a specific, unusual, type of
jet that contains one very hard parton and is very narrow; selecting
(`triggering on') hadrons therefore constitutes selecting an unusual
sample of jets that lose less energy, and this selection effect becomes
stronger at higher $p_{T}$. This is one reason that $R_{{\rm AA}}$
for hadrons rises at the highest $p_{T}$ even though $R_{{\rm AA}}$
for jets remains comparably suppressed. 

The second reason why some jets lose more energy than others is that
when two or more jets are produced in a collision they each traverse
different lengths of QGP. There is evidence for this effect in measurements
of a $v_{2}$-like anisotropy for particles with high transverse momentum
that originate in jets \cite{Sirunyan:2017pan}: these jets typically
lose less energy when moving along the short axis, as measured by
the event-plane angle of the $v_{2}$ at low $p_{T}$ (described in
Section \ref{sec:A-hydrodynamic-fluid}). All of this is to say that
parton energy loss is a dominant effect contributing to the modification
of many jet observables in heavy ion collisions as compared to proton-proton
collisions. $dE/dx$, the rate of parton energy loss in plasma, is
parametrized in different ways for partons that are assumed to be
traversing a weakly coupled plasma versus for those which are assumed
to be traversing a strongly coupled plasma that behaves as it would
in a holographic gauge theory \cite{Chesler:2014jva,Chesler:2015nqz,Casalderrey-Solana:2014bpa,Casalderrey-Solana:2015vaa}.
In either case, present data is being used to constrain the magnitude
of $dE/dx$ and in the near-future, as the precision of the data improves
further, it should become possible to differentiate between different
choices for the $T$-, $x$- and $E$-dependence of $dE/dx$.

The energy and momentum `lost' by a jet in a heavy ion collision is,
of course, not lost. We now know from experiment that it ends up shared
among many soft hadrons in the final state of the collision that are
spread out over a wide range of angles, up to 60 or even 120 degrees,
around the jet direction \cite{Khachatryan:2015lha,CMSsoft}. This
is certainly qualitatively consistent with a picture in which the
jet excites a wake in the droplet of QGP, namely a region of moving
and perhaps heated plasma behind the jet that carries the momentum
in the jet direction `lost' by the jet. Like the unperturbed plasma,
this wake becomes many soft hadrons after the droplet of QGP falls
apart into hadrons. Because they carry net momentum in the jet direction,
some of the hadrons from the wake must end up within what experimentalists
see as the jet \cite{Casalderrey-Solana:2016jvj,Milhano:2017nzm}.
This means that a quantitative understanding of the wake is a prerequisite
to a quantitative understanding of the soft component of jets reconstructed
in heavy ion collisions. Quantitative studies of the hydrodynamics
of these wakes are now being done \cite{Chen:2017zte} and theorists
should soon be able to do large-scale Monte Carlo calculations which
track jet production in a hard scattering, jet showering, jet quenching,
and the hydrodynamics of the specific wake produced by each specific
jet (see i.e. \cite{Cao:2017zih}). Although full-scale calculations
remain to be done, there are preliminary indications in some calculations
\cite{Casalderrey-Solana:2016jvj,Hulcher:2017cpt} that the wakes
made by jets shooting through the plasma do not have time to fully
hydrodynamize, as they yield more 2-4 GeV hadrons and fewer 0-2 GeV
hadrons than they would have if they had completely hydrodynamized
\cite{Casalderrey-Solana:2016jvj}. This is exciting as it raises
the prospect of using jets, specifically their wakes, to obtain experimental
access to the physics of hydrodynamization. In this way, analysis
of jets in heavy ion collisions may yield insights into how QGP forms
as a function of time in addition to, in the longer term, revealing
how QGP emerges as we coarsen the resolution scale of the microscope
with which we probe it. Achieving this longer term goal will require
having a quantitative understanding built upon precise data of energy
loss (which results in an ensemble of narrower jets), jet wakes (which
make jets as observed wider), and the accumulation of transverse momentum
by the jet partons via their soft interactions with the liquid QGP.
Only then will it be possible to look for the rare (power-law-rare,
not exponentially rare) hard scattering events in which a parton within
a jet (or, even more rarely, a jet itself) gets kicked by a detectable
angle as it resolves, and scatters off, a parton within the liquid. 

It is at present an unfortunate aspect in the studies of hard probes
that experimentally only the final particles can be measured: it is
in general not possible to directly compare a probe before and after
passing through QGP. Recently this situation has been improved by
selecting events with an energetic photon or Z-boson and one or more
energetic jets \cite{Chatrchyan:2014csa}, which have the advantage
that the photon or Z-boson is unperturbed by the plasma and hence
gives some probabilistic information about the energy of the jet or
jets produced in the same event. Nevertheless, measuring the photon
yields little information about the width of the jet, which plays
an important and perhaps dominant role in determining how much energy
it loses. In this context, it is exciting that experimentalists have
recently begun to measure a host of different jet substructure observables,
beyond the traditional jet shape, jet width, jet fragmentation function
and jet mass, that are constructed in a variety of different ways
via grooming jets and obtaining operational measures of their substructure
\cite{Sirunyan:2017bsd,Acharya:2017goa,Kauder:2017mhg}. Although
this has not yet been realized, it may be possible to identify an
observable that is (relatively) unmodified by the passage of a jet
through QGP and that in $pp$ collisions is in one-to-one correspondence
with the width of the jet. If this potential is realized, by measuring
other observables that are sensitive to energy loss as a function
of this observable it will be possible to study the quenching of jets
for which we have some information about what their widths would have
been in the absence of quenching.

\section{Summary and big questions\label{sec:Big-questions}}

\begin{summary}[SUMMARY POINTS]
\begin{enumerate}
\item We study heavy ion collision to gain insight into perhaps the simplest
form of complex matter, described by the fundamental laws of QCD.
This super hot liquid filled the microseconds old universe, making
it the first complex matter to form as well as the source of all protons
and neutrons. Heavy ion collisions are little bangs, recreating droplets
of big bang matter.
\item Within a time of order of 1 fm/c, the matter and entropy produced
in a heavy ion collision form a droplet of strongly coupled QGP, evolving
according to relativistic hydrodynamics with very small specific viscosity.
\item QGP is neither a collection of hadrons nor a nearly free gas of quarks
and gluons. The colored quarks are free to diffuse and are not confined,
but at the same time they are always very strongly coupled with their
neighbors in the liquid.
\item Hydrodynamics converts spatial anisotropies into momentum anisotropy,
giving us a direct experimental probe of both the spatial geometry,
which is the source of the anisotropies, and the viscosity, which
seeks to dissipate them. The QGP is very lumpy when it forms. As it
expands and cools hydrodynamically, as the lumps smooth out the resulting
momentum anisotropies persist because the specific viscosity of QGP
is small.
\item The strongly coupled nature of QGP is also seen and probed at a broad
range of length scales by jet quenching: the rapid loss of energy
by highly energetic partons traversing QGP.
\item There is a wealth of experimental data, from longitudinal rapidity
and transverse momentum distributions to quark flavor, and from two
particle to multiparticle correlations, which are surprisingly similar
across a variety of colliding systems spanning three orders of magnitude
in both volume and energy.
\end{enumerate}
\end{summary}

\begin{issues}[BIG QUESTIONS]
\begin{enumerate}
\item How does QGP form and hydrodynamize within 1 fm/c? What are the qualitative
differences, if any, between the description of hydrodynamization
in a heavy ion collision obtained by assuming a weakly coupled initial
stage versus a strongly coupled holographic calculation? Note that
perturbative calculations typically treat $\alpha_{s}=0.3$ as small
while holographic calculations treat the corresponding 't Hooft coupling
$\lambda\approx11$ as large. What can we learn about the timescales
and dynamics of hydrodynamization, and hence QGP formation, by analyzing
the wakes that jets leave behind as they traverse a droplet of QGP?
\item What are the limits of the applicability of hydrodynamics? Can it
be applied even to systems of size a fermi or less? What is the smallest
droplet of QGP that behaves hydrodynamically, and how does the answer
to this question change at very high temperatures where $\eta/s>1$
and QGP is no longer a strongly coupled liquid?
\item How does a strongly coupled liquid emerge when QGP is analyzed with
a spatial resolution of order $1/T$ or coarser, given that because
QCD is asymptotically free what you will see at much finer resolution
is weakly coupled quarks and gluons? How can we use jets to see the
inner workings of QGP and answer this question? If we can understand
how QGP emerges from an asymptotically free gauge theory, can we use
this understanding to obtain general lessons about how complex forms
of matter emerge from simple underlying laws? 
\item How can we relate measurements of the gluon distribution in nuclei
made at a future Electron Ion Collider to the distribution of the
energy density across the transverse plane immediately after a heavy
ion collision \textemdash{} quantitatively?
\item Can we obtain an experimental determination, even indirectly, of the
temperature of the matter produced in a heavy ion collision at a time
at which we can also determine its energy density? If we could, we
could obtain an experimental determination of the number of thermodynamic
degrees of freedom, the quantity whose increase reflects the liberation
of color above the crossover in the QCD phase diagram. 
\item How do the hydrodynamics of QGP and the thermodynamics of its transition
to hadronic matter as it cools change as QGP is doped with an excess
of quarks over antiquarks? Is there a critical point in the region
of the QCD phase diagram that heavy ion collisions can explore, or
do all collisions that make QGP only explore a crossover in the phase
diagram?
\item Can we explain the distribution of energy and entropy (particle multiplicity)
as a function of rapidity in heavy ion collisions over a wide range
of collision energies from first principle computations? Ditto for
hadronization, and in particular can we explain why hadronization
produces hadrons in chemical equilibrium? More generally, why are
many bulk phenomena so similar for $AA$, $pA$, $pp$, $\pi{\rm A}$
and in some cases even $e^{+}e^{-}$ collisions, over an enormous
range of collision energies?
\item Is there color superconducting quark matter at the centers of some
or all neutron stars? This important question about the phase diagram
of QCD cannot be addressed by heavy ion collisions; we hope that observations
of binary neutron stars colliding and merging will help.
\end{enumerate}
\end{issues}

\section*{Acknowledgments}

We are pleased to acknowledge helpful comments from Gian Michele Innocenti,
Guilherme Milhano, Greg Ridgway, Raju Venugopalan, Jing Wang, Ryan
Weller and Bill Zajc. This work is supported by the U.S. Department
of Energy under grant Contract Number DE-SC0011090. WS is supported
by VENI grant 680-47-458 from the Netherlands Organisation for Scientific
Research (NWO).

\bibliographystyle{bibstyle}
\bibliography{references}

\end{document}